\documentclass[12pt]{article}
\usepackage[utf8]{inputenc}
\usepackage[british]{babel}
\usepackage{cmap}
\usepackage{lmodern}

\usepackage{amssymb, amsmath, amsthm}
\usepackage[a4paper,top=25mm,bottom=25mm,left=25mm,right=25mm]{geometry}
\usepackage{ragged2e}

\usepackage{authblk} 
\usepackage{pifont}
\usepackage{graphicx}
\usepackage[dvipsnames,svgnames,table]{xcolor}
\usepackage[figuresright]{rotating}
\usepackage{xtab} 
\usepackage{longtable} 
\usepackage{multirow}
\usepackage{footnote}
\usepackage[stable]{footmisc}
\usepackage{chngpage} 
\usepackage{pdflscape} 
\usepackage[nottoc,notlot,notlof]{tocbibind} 

\usepackage{pgfplots}
\pgfplotsset{every tick label/.append style={font=\footnotesize}}
\pgfplotsset{compat=1.18}
\usepackage{setspace}

\usepackage{array}
\newcolumntype{K}[1]{>{\centering\arraybackslash$}p{#1}<{$}}

\makesavenoteenv{tabular}
\usepackage{tabularx}
\usepackage{booktabs}
\usepackage{threeparttable}
\usepackage[referable]{threeparttablex} 
\newcolumntype{R}{>{\raggedleft\arraybackslash}X}
\newcolumntype{L}{>{\raggedright\arraybackslash}X}
\newcolumntype{C}{>{\centering\arraybackslash}X}
\newcolumntype{A}{>{\columncolor{gray!25}}C}
\newcolumntype{a}{>{\columncolor{gray!25}}c}

\newlength{\tablen}

\usepackage{dcolumn} 
\newcolumntype{.}{D{.}{.}{-1}}

\usepackage{tikz}
\usetikzlibrary{arrows, calc, matrix, patterns, positioning, trees}
\usepackage[semicolon]{natbib} 
\usepackage[hyphens]{xurl}
\usepackage{microtype}
\usepackage[justification=centering]{caption} 

\usepackage[labelformat=simple]{subcaption}

\DeclareCaptionLabelFormat{parenthesis}{(#2)}
\makeatletter
\renewcommand\p@subfigure{\arabic{figure}.}
\makeatother

\DeclareCaptionLabelFormat{parenthesis}{(#2)}
\makeatletter
\renewcommand\p@subtable{\arabic{table}.}
\makeatother

\usepackage{enumitem}

\setlist[itemize]{leftmargin=2.5\parindent}
\setlist[enumerate]{leftmargin=2.5\parindent}

\usepackage{hyperref} 
\hypersetup{
  colorlinks   = true,    		
  urlcolor     = blue,    		
  linkcolor    = blue,    		
  citecolor    = ForestGreen	
}

\def\addlegendimage{\csname pgfplots@addlegendimage\endcsname}

\theoremstyle{plain}

\theoremstyle{definition}

\theoremstyle{remark}

\makeatletter
\let\@fnsymbol\@alph
\makeatother

\def\keywords{\vspace{.5em} 
{\noindent \textit{Keywords}: }}

\def\AMS{\vspace{.5em} 
{\noindent \textbf{\emph{MSC} class}: }}

\def\JEL{\vspace{.5em} 
{\noindent \textbf{\emph{JEL} classification number}: }}

\title{The uncertainty of a tournament draw: \\ Insights from the Champions League}

\author{
\href{https://sites.google.com/view/laszlocsato}{L\'aszló Csat\'o}\thanks{~Institute for Computer Science and Control (SZTAKI), Hungarian Research Network (HUN-REN), Laboratory on Engineering and Management Intelligence, Research Group of Operations Research and Decision Systems, Budapest, Hungary \newline
Corvinus University of Budapest (BCE), Institute of Operations and Decision Sciences, Department of Operations Research and Actuarial Sciences, Hungary \newline
\texttt{laszlo.csato@sztaki.hun-ren.hu}} \qquad
Andr\'as Gyimesi\thanks{~University of P\'ecs, Hungary \newline
Institute for Computer Science and Control (SZTAKI), Hungarian Research Network (HUN-REN), Laboratory on Engineering and Management Intelligence, Research Group of Operations Research and Decision Systems, Budapest, Hungary \newline
\texttt{gyimesi.andras@ktk.pte.hu}} \qquad
\href{https://research.ugent.be/web/person/dries-goossens-0/en}{Dries Goossens}\thanks{~Ghent University, Department of Business Informatics and Operations Management, Belgium \newline
FlandersMake@UGent -- core lab CVAMO, Ghent, Belgium \newline
\texttt{dries.goossens@ugent.be}} \\
Karel Devriesere\thanks{~Ghent University, Department of Business Informatics and Operations Management, Belgium \newline
FlandersMake@UGent -- core lab CVAMO, Ghent, Belgium \newline
\texttt{karel.devriesere@ugent.be}}  \qquad 
Roel Lambers\thanks{~Alliander, The Netherlands \newline
\texttt{lambers.roel@gmail.com}} \qquad
\href{https://feb.kuleuven.be/public/u0037710/}{Frits Spieksma}\thanks{~Eindhoven University of Technology, Department of Mathematics and Computer Science, The Netherlands \newline
\texttt{f.c.r.spieksma@tue.nl}}
}
\date{\today}

\begin{document}
\newgeometry{top=15mm,bottom=15mm,left=25mm,right=25mm}

\maketitle
\thispagestyle{empty}

\begin{abstract}
\noindent
The group draw of major sports tournaments implies some uncertainty, with lucky teams often enjoying a substantial unfair advantage. For the first time in the literature, we propose a technique to quantify this draw uncertainty, which, arguably, should be minimised. Our simulation-based approach requires generating a representative set of random draws to compute the variance of qualifying probabilities for each team.
The method is applied to compare draw uncertainty in the former group stage and the current incomplete round-robin league phase of the UEFA Champions League, under both accurate and inaccurate seedings. We also break down the impact of the 2024/25 reform into various components.
The new format is found to decrease draw uncertainty, but the reduction is mainly attributable to the inaccurate seeding system used by UEFA. Consequently, the primary benefit of an incomplete round-robin tournament compared to the standard group stage lies in the robustness of its draw uncertainty to the seeding of the teams.

\end{abstract}

\keywords{OR in sports; seeding; simulation; tournament design; UEFA Champions League}
 
\AMS{62P20, 90-10, 90B90}

\JEL{C44, C53, Z20}

\clearpage
\restoregeometry

\section{Introduction}

A sports tournament can only be engaging when there is an \emph{a priori} uncertainty about its winner. The literature has primarily focused on \emph{outcome uncertainty} \citep{CoatesHumphreysZhou2014, ScarfParmaMcHale2019, BuraimoForrestMcHaleTena2022}, referring to the unpredictability of individual match results as team abilities do not perfectly determine outcomes. However, uncertainty may also pertain to the set of matches themselves if the tournament structure involves a random draw procedure. We call this \emph{draw uncertainty}. 

While outcome uncertainty is generally regarded as an intrinsic and desirable feature of competitive sport, draw uncertainty is an exogenous element of chance unrelated to athletic performance and entirely caused by the organiser. Consequently, unlike match uncertainty, draw uncertainty should always be minimised. Indeed, draw uncertainty is absent in certain popular tournament formats, such as knockout competitions with fixed brackets (used, for example, in the second stage of the FIFA World Cup) and double round-robin leagues (the common format in European sports competitions). Moreover, when draw uncertainty is present, policy-makers often implement sophisticated seeding mechanisms to prevent the top teams from facing each other in the early stages of the competition \citep{ScarfYusofBilbao2009, DevriesereCsatoGoossens2025}. For instance, the first stage of the FIFA World Cup aims to reduce the imbalance across the groups \citep{CeaDuranGuajardoSureSiebertZamorano2020,  CsatoBeckerDevriesereGoossens2026, Guyon2015a, LalienaLopez2019, LalienaLopez2025}.

Therefore, decision makers could benefit from measuring draw uncertainty, especially to evaluate the effect of various rule changes in tournament design, which are quite common in professional sports \citep{KendallLenten2017}.
Our paper addresses this research gap. We propose an innovative approach to quantify draw uncertainty for each team by the variance in their probability of obtaining the prize under a representative set of draw outcomes. While the measure might seem straightforward at first sight, its computation usually requires combining state-of-the-art techniques from operational research. First, generating a high number of valid draws may be challenging, as illustrated by the draws of the FIFA World Cup group stage \citep{Guyon2015a, RobertsRosenthal2024, Csato2025c, CsatoBeckerDevriesereGoossens2026, Csato2027} and the UEFA Champions League league phase \citep{DevriesereGoossensSpieksma2026, GuyonBenSalemBuchholtzerTanre2026}. Second, the results of individual matches need to be simulated by an appropriate prediction model \citep{ScarfYusofBilbao2009, ChaterArrondelGayantLaslier2021}, for which purpose we adopt the independent Poisson model of \citet{Maher1982}.

The suggested measure of draw uncertainty is applied to the UEFA Champions League.
This high-profile tournament provides an ideal case study due to a recent fundamental rule change. Between the 2003/04 and 2023/24 seasons, 32 teams played in eight groups such that in each group, four teams contested qualification for the Round of 16 in a double round-robin format. Since the 2024/25 season, 36 teams compete in an incomplete round-robin league phase where they play against eight different opponents. The top eight teams directly qualify for the Round of 16, and the teams ranked from 9th to 24th play against each other in the newly introduced knockout phase play-offs to reach the Round of 16.
As the 2024/25 reform consists of several components (see Section~\ref{Sec41}), a decomposition method is developed to disentangle, for each team separately, the effects of
(a) the inaccurate seeding;
(b) the knockout phase play-offs; and
(c) the change from the multi-group structure to a single incomplete round-robin league.

Draw uncertainty can be interpreted not only at the level of individual teams, but also at the level of a tournament, which makes it possible to compare alternative tournament designs with respect to their draw uncertainty. This requires aggregating the draw uncertainties faced by the teams into a tournament-specific measure of draw uncertainty; the average, median, or maximum seem to be the most straightforward options. In the following, we do not deal with this aggregation procedure because, in our case study of the UEFA Champions League, draw uncertainty generally changes in the same direction for all teams, albeit the magnitude of these changes differs.

Crucially, we find that the main advantage of the incomplete round-robin format compared to the traditional group stage is the substantially lower (more favourable) draw uncertainty if the seeding and team strengths are not aligned. In particular, if the seeding is determined by UEFA club coefficients, while team strengths are based on Football Club Elo Ratings, using an incomplete round-robin format substantially reduces draw uncertainty compared to the group stage format. This finding is especially relevant to UEFA since their official seeding policy is known to be inaccurate \citep{Csato2024c}. Conversely, if the seeding and team strengths are fully aligned, the draw uncertainty of the incomplete round-robin league and the traditional group stage remains about the same. 
Finally, relatively weak teams benefit less from a lucky assignment in an incomplete round-robin tournament than they do in the traditional group stage, thereby leading to a fairer design.

These results uncover the benefits of using an incomplete round-robin tournament and clearly show what insights can be gained by applying operational research to tournament design problems. According to our findings, the recent fundamental rule change in the UEFA Champions League has not had unwanted consequences, at least with respect to draw uncertainty. This is an important message because many reforms aimed to make a sports tournament more exciting have had undesirable effects \citep{KendallLenten2017}.

The paper is organised as follows.
Section~\ref{Sec2} presents how this study is connected to the existing literature. Section~\ref{Sec3} introduces and motivates our measure of draw uncertainty. The 2024/25 reform in the UEFA Champions League is discussed in Section~\ref{Sec4}. Section~\ref{Sec5} shows and explains the results, and Section~\ref{Sec6} concludes.

\section{Related literature} \label{Sec2}

The current paper is related to at least three research areas.
First, a number of works have addressed tournament design issues in the history of the UEFA Champions League, often via simulations.
\citet{ScarfYusofBilbao2009} compare several tournament designs of this competition to determine the value of various tournament metrics such as the proportion of unimportant games and the average rank of the winner.
The impact of the reform in the seeding system in 2015/16 (Section~\ref{Sec413}) is analysed by \citet{CoronaForrestTenaHorrilloWiper2019} and \citet{DagaevRudyak2019}.
\citet{Csato2022b} evaluates the effect of a substantial change in the Champions Path of the UEFA Champions League qualification system in 2018/19.
\citet{CsatoMolontayPinter2024} compute the probability of a stakeless match (when the rank of a team does not depend on the outcome of the match) in the UEFA Champions League group stage under all reasonable schedules.
\citet{CabralLiao2025} show that access to the UEFA Champions League is \emph{not} based on the idea of cross-league fairness: the marginally excluded team from top domestic leagues is expected to outperform marginally included teams from several lower-ranked leagues.

Second, our results contribute to the understanding of incomplete round-robin tournaments.
Balancing the strength of opponents in such tournaments has been extensively investigated, see \citet{FreybergKeranen2023, Froncek2013, FroncekShepanik2016, FroncekShepanik2018, FroncekShepanik2022}. \citet{LiVanBulckGoossens2025} propose the incomplete round-robin format to organise multi-league sports competitions: its flexibility is exploited to decrease total travel distance and venue capacity violations. A metaheuristic based on Benders’ decomposition is developed and validated using real-world benchmarks. The advantage of an incomplete round-robin format compared to the traditional structure with round-robin groups is verified.
\citet{DevriesereGoossens2025} describe how Belgian field hockey youth competitions can be redesigned as an incomplete round-robin tournament. This novel approach is able to decrease total travel time by up to 25\% compared to the official schedule, and can provide a solution where 94\% of the teams are better off with a reduction of 20\% in total travel time. Convinced by these gains, the Belgian Royal Hockey Association has adopted their proposal.

Some recent papers study the incomplete round-robin format of the UEFA Champions League.
\citet{Gyimesi2024} quantifies short-, mid-, and long-term competitive balance in the old and new designs. The reform improves competitive balance, especially regarding match uncertainty and the occurrence of stakeless matches.
\citet{DevriesereGoossensSpieksma2026} estimate the effect of the new format on the expected number of three types of non-competitive matches: asymmetric (when exactly one team is indifferent), stakeless (when both teams are indifferent), and collusive (when no team is indifferent but both can achieve their desired prize with a particular outcome). 
\citet{CsatoGyimesi2026b} use a probabilistic model to classify the matches played in the last round of the UEFA Champions League group stage and league phase.
\citet{WinkelmannDeutscherMichels2026} aim to give guidance on the number of points required to qualify by accounting for the low proportion of draws observed in the 2024/25 UEFA Champions League. The results are compared to the more competitive UEFA Europa League, which is organised in the same format since the 2024/25 season. The probability of draws turns out to substantially affect qualification probabilities around the critical thresholds.
\citet{CsatoWinkelmann2026} demonstrate that the novel incomplete round-robin format creates stronger incentives for offensive play compared to the previous group stage, which might explain the reduced probability of draws.
\citet{GuyonBenSalemBuchholtzerTanre2026} discuss the draw procedure used in the incomplete round-robin league phase. Four reasonable draw methods, including the official UEFA mechanism, are compared with respect to their fairness (the distributions of average opponent strength for the 36 teams) and matching probabilities. An interactive simulator is available at \url{https://julienguyon.github.io/UEFA-league-phase-draw}.
\citet{HautBoeyRigaut2026} investigate the impact of UEFA Champions League format changes via simulations by focusing on competition outcomes, the financial threshold for success, and the distribution of matches between high-profile and lower-ranked clubs.

Third, academic research has also focused on \emph{balance}: the groups should be at the same competitive level to avoid a situation when a weak team playing against weak opponents has a higher chance to qualify than a strong team playing against strong opponents. \citet{Guyon2015a} has proved that the 2014 FIFA World Cup draw produced unbalanced groups, which has prompted a subsequent change in the draw procedure \citep{Guyon2018d}. Nevertheless, the 2022 FIFA World Cup draw has again failed to guarantee balance \citep{Csato2023d}, and finding a draw system that creates balanced groups remains a popular topic in operations research \citep{CeaDuranGuajardoSureSiebertZamorano2020, LalienaLopez2019, LalienaLopez2025}. There are empirical and simulation studies on the effect of the group draw, too.
\citet{LaprePalazzolo2022}, \citet{LaprePalazzolo2023}, and \citet{LapreAmato2025} use logistic regressions to quantify the impact of imbalanced groups on the probability of success in the FIFA Women's World Cup (between 1991 and 2019), the FIFA Men's World Cup (between 1954 and 2022), and the UEFA European Championship (between 1980 and 2024), respectively.
\citet{Avila-CanoTriguero-Ruiz2024} find that, even though the UEFA Champions League groups were not homogeneous with respect to ex ante and ex post competitive balance, their composition had no effect on which team would be the champion.
\citet{Csato2025c} assesses via simulations how the distortions of the 2018 FIFA World Cup draw procedure have changed the probability of qualifying for the knockout stage.

However, we are not aware of any study explicitly investigating draw uncertainty.

\section{Measuring draw uncertainty} \label{Sec3}

Consider a tournament where the contestants compete for a given prize. If each contestant faces each other contestant the same number of times (i.e., a round-robin tournament), the probability of obtaining the prize depends only on the relative strengths of the contestants. However, in many tournaments, there are too many contestants or too few time slots to play a (complete) round-robin tournament. In this case, the contestants face only a subset of their possible opponents, and their probability of success is also influenced by the strength of the opponents they face. Typically, this subset of opponents is determined by a draw, and the draw itself adds to the uncertainty of the tournament outcome. In fact, draw uncertainty can make a competition unfair by threatening the principle of equal treatment of equals, and systematically favour a team whose opponents are weaker than the opponents of another team of the same strength. This has been demonstrated recently for both the FIFA Men's \citep{LaprePalazzolo2023} and Women's World Cup \citep{LaprePalazzolo2022}, as well as for the UEFA European Championship \citep{LapreAmato2025}.

We propose to quantify draw uncertainty, for each team, as the \emph{standard deviation} of the probabilities of obtaining the prize over multiple possible draw outcomes. Indeed, if these probabilities for a team are roughly equal over all draws, the standard deviation remains low, reflecting that the impact of the draw is small for this particular team. However, if the qualifying probabilities for a team vary significantly over the draws, the standard deviation will be high, indicating that the draw adds a lot of uncertainty to the tournament outcome.

Unfortunately, a theoretical analysis of this measure is heavily complicated due to the high number of possible outcomes, even if there are only eight contestants \citep{McGarrySchutz1997}. Therefore, we use Monte Carlo computer simulations for real-world tournaments. In particular, we suggest generating $K$ random draws and $M$ random sets of match outcomes (called a \emph{scenario}) for each of these draws. Given a draw and a team, the qualifying probability of the team is given by the relative frequency of the scenarios in which the team obtains the prize. Finally, the standard deviation $\sigma_i$ of these $K$ qualifying probabilities (one for each draw) measures draw uncertainty for a given team $i$.


A mathematically rigorous interpretation of $\sigma_i$ on its own is challenging. However, it is useful to evaluate the impact of any rule change by comparing the standard deviation of the draw in the new design ($\sigma_i^n$) to the standard deviation of the draw in the old design ($\sigma_i^o$). This comparison reveals the change in the impact of the draw on securing the prize.

Another issue that we can explore with this approach is the effect of inaccurate seeding. Seeding is the ordering of the contestants according to their strength (or merit) before the tournament, typically based on historical performances and/or the judgement of experts \citep{ScarfYusof2011}. However, actual team strengths are never known exactly, and tournament organisers should always keep in mind that the seeding underlying the draw can be inaccurate: some teams might be substantially stronger (or weaker) than assumed at the time of the draw. Consequently, an ideal tournament should not only have a low level of draw uncertainty, but it also needs to be \emph{robust} with respect to the seeding; that is, draw uncertainty should not increase significantly if the seeding becomes inaccurate.

The assumption of one prize may be unrealistic: contestants in a tournament might have different expectations and, consequently, different prizes in mind (e.g.\ winning the tournament, or qualifying for the next round). However, the proposed measure could equally be applied to other prizes, allowing draw uncertainty to be assessed from various perspectives. The relevant prize can even differ across the teams.

\section{A case study: the 2024/25 reform of the UEFA Champions League} \label{Sec4}

Section~\ref{Sec41} details the changes in the design of the UEFA Champions League. Section~\ref{Sec42} discusses how we simulate the old and the new tournament designs as closely as possible to the UEFA rules. The simulation model for generating match outcomes is described in Section~\ref{Sec43}. Section~\ref{Sec44} adopts the draw uncertainty measure suggested in Section~\ref{Sec3} to our case study in order to decompose the effects of the 2024/25 reform.

\subsection{Tournament designs of the UEFA Champions League} \label{Sec41}

We first present the differences between the two competition designs called \emph{old} (used until the 2023/24 season) and \emph{new} (used from the 2024/25 season), following the logic of a recent survey by \citet{DevriesereCsatoGoossens2025}.
This description allows us to measure draw uncertainty (according to the ideas discussed in Section~\ref{Sec3}), thereby enabling an analysis of the effect of the fundamental reform.


\subsubsection{Participating teams} \label{Sec411}

The old format involved 32 teams. The ranking based on UEFA association coefficients determined the number of participating teams for each association; the four highest-ranked associations provided four teams each. Lower-ranked nations had fewer participating teams, and several champions had to play one or more qualification rounds to enter the tournament.

The new format has retained the same qualification system; however, four additional spots have been allocated. First, the fifth-ranked association has received an extra direct slot. Second, the number of teams qualifying via the Champions Path has increased by one. Third, one additional spot is given to each of the two associations with the best collective performance in the previous season of UEFA club competitions \citep{CsatoIlyin2025}. Since the 2003/04 season, these leagues are usually two from the four leading leagues (England, Germany, Italy, Spain); the only exceptions were France (2003/04), Romania (2005/06), Ukraine (2008/09), Portugal (2010/11), and the Netherlands (2021/22).


\subsubsection{Tournament format} \label{Sec412}

Between the seasons 2003/04 and 2023/24, the format of the UEFA Champions League did not change. The tournament started with a group stage of 32 teams that played in eight groups of four teams each. The groups were organised in a double round-robin format, that is, each team played against all the others once at home and once away. The group winners and runners-up qualified for the Round of 16, the third-placed teams were transferred to the second most prestigious UEFA club football competition (called UEFA Europa League since 2008/09), while the last teams were eliminated. In the Round of 16, group winners were matched with the runners-up subject to some further constraints \citep{KlossnerBecker2013}.

From the 2024/25 season onwards, the group stage has been replaced by the incomplete round-robin league phase, which is contested by all the 36 teams. Each team plays eight matches, four at home and four away. Then, a ranking of the 36 teams is constructed based on the number of points each team collected in these eight matches. The first eight teams directly qualify for the Round of 16, the next 16 teams play against each other in the play-offs. In particular, the teams form four seeded pairs (teams in positions 9 and 10, 11 and 12, 13 and 14, and 15 and 16) and four unseeded pairs (positions 17 and 18, 19 and 20, 21 and 22, and 23 and 24). The teams in the $k$th seeded pair play against the teams in the $(5-k)$th unseeded pair ($k=1,2,3,4$); for example, the 11th-ranked team (as well as the 12th-ranked team) has an equal (50\%) chance to play against either the team ranked 21st or 22nd. The same principle applies to the pairing in the Round of 16.
The last 12 teams (ranked from 25th to 36th) are eliminated.


\subsubsection{Seeding} \label{Sec413}

The seeding regime of the old design was reformed in the 2015/16 and the 2018/19 seasons \citep{Csato2020a}. Until the 2014/15 season, the teams were assigned to four pots based on their UEFA club coefficients established at the beginning of the season. Pot 1 contained the titleholder and the seven highest-ranked teams, while Pots 2, 3, 4 consisted of the other teams according to their ranking order.
The 2015/16 reform \citep{CoronaForrestTenaHorrilloWiper2019, DagaevRudyak2019} placed the titleholder and the champions of the seven highest-ranked associations in Pot 1. Between 2018/19 and 2023/24, Pot 1 contained the UEFA Champions League and Europa League titleholders together with the champions of the six best associations. Vacant slots were filled by the champions of the seventh- and eighth-ranked associations, if necessary.

From the 2024/25 season onwards, UEFA has reinstated the original seeding policy used until the 2014/15 season. In the league phase, the 36 teams are seeded into four pots of nine teams each based on their UEFA club coefficients. The only exception is that the Champions League titleholder is automatically assigned to the first pot.


\subsubsection{Draw} \label{Sec414}

In the old design, each group contained one team from each of the four pots. Teams from the same association could not be drawn into the same group. In addition, UEFA formed pairs of teams from the same nation to guarantee that these teams play on different days in Groups A--D and E--H, respectively. These TV pairings do influence the draw probabilities \citep{Guyon2021a}.

In the new format, each team plays against two different opponents from each pot, including its own pot. One of these matches is played at home, and the other is played away. Teams from the same association cannot play against each other, and no team can play against more than two teams from the same association \citep{UEFA2024f}.

As an aside, ensuring that all matches can be played within eight matchdays requires the draw to satisfy a graph-theoretic condition. Consider the graph whose vertices represent the teams and whose edges represent the matches determined by the draw. This is an 8-regular graph and has an edge chromatic number of either 8 or 9 by Vizing's theorem \citep{Vizing1964}. Only draws that correspond to an 8-edge-colorable graph are feasible, since a graph with edge chromatic number 9 would require nine matchdays to schedule all matches. The probability of obtaining a 9-edge-colorable graph in the draw is positive \citep{GuyonBenSalemBuchholtzerTanre2026}, albeit extremely low.




\subsubsection{Ranking} \label{Sec415}

The teams are ranked according to the number of points collected: 3 points for a win, 1 point for a draw, and 0 points for a loss.
In the old design, the tie-breaking rules were head-to-head results (number of points, goal difference, goals scored in all matches among the tied teams), applied recursively if necessary. The remaining ties were decided by goal difference and the number of goals scored. 

In the new design, the first tie-breaking criterion is goal difference, followed by goals scored, away goals scored, wins, away wins, number of points obtained collectively by the opponents, collective goal difference of the opponents, number of goals scored by the opponents, disciplinary record, and UEFA club coefficient. In the 2024/25 season, the order of Real Madrid and Bayern M\"unchen has been decided by more away wins for the Spanish club.

\subsubsection{Overview of the 2024/25 reform} \label{Sec416}

The main changes between the old and the new UEFA Champions League designs can be summarised as follows:
\begin{itemize}
\item
The number of teams has increased from 32 to 36;
\item
The format of playing a double round-robin in eight groups of four teams has been replaced by 36 teams playing in a single incomplete round-robin;
\item 
The number of matches played by a team has increased from six to eight, with one additional home and one additional away match against two different teams from the own pot (supposed to be of comparable strength);
\item
A play-off round has been introduced for the teams placed 9th to 24th to determine qualification to the Round of 16;
\item 
The seeding has removed any preference given to the champions of the strongest associations.
\end{itemize}

\subsection{Simulating the old and new Champions League designs} \label{Sec42}


The new design contains four additional teams compared to the old design. The allocation of the two European performance spots is straightforward since they are given to the top two associations in the previous season of UEFA club competitions \citep{CsatoIlyin2025}. In contrast, the extra slot for the fifth-ranked association in the new design affects a team that qualified in some seasons via the League Path of the qualification even in the old design. In these cases, this extra slot is given to the team eliminated by the smallest margin in the League Path of the qualification. Finally, the additional slot given to the Champions Path of the qualification in the new design is assumed to be obtained by the team eliminated by the smallest margin in the Champions Path of the qualification. The same procedure is applied in a reversed direction to remove teams from the new design; that is, the fourth-placed team of the fifth-ranked association and the team qualifying by the smallest margin in the Champions Path are not considered in the old design.

\begin{table}[t!]
  \centering
  \caption{Identification of the four additional teams in the new Champions League design}
  \label{Table1}
\centerline{
\begin{threeparttable}
  \rowcolors{3}{}{gray!20}
    \begin{tabularx}{1\textwidth}{l LLll} \toprule
    Season & EPS 1 & EPS 2 & Champions Path & League Path \\ \bottomrule
    2019/20 & Arsenal & Getafe & Young Boys & LASK \\
    2020/21 & Villarreal & Bayer Leverkusen & Molde & PAOK \\
    2021/22 & Leicester City & Real Sociedad & Ludogorets & Monaco \\
    2022/23 & Arsenal & PSV Eindhoven & Crvena Zvezda & Monaco \\
    2023/24 & Liverpool & Atalanta & Rakow Czestochowa & Marseille \\ \hline
    2024/25 & Bologna & Dortmund & Slovan Bratislava & Lille \\
    2025/26 & Newcastle United & Villarreal & Kairat & Monaco \\ \toprule
    \end{tabularx}
\begin{tablenotes} \footnotesize
\item
EPS 1 (2) is given to the association with the highest (second-highest) yearly country coefficient (the number of points divided by the number of participating teams) in the previous season.
\item
In the Champions Path, the team that was eliminated (qualified) by the smallest margin is added to (removed from) the new design.
\item
In the League Path, the third-placed (fourth-placed) team of the fifth-ranked association is added to (removed from) the new design. If this team qualified in the old design, the team that was eliminated by the smallest margin is added to the new design.
\end{tablenotes}
\end{threeparttable}
}
\end{table}

Table~\ref{Table1} identifies these teams in all seasons between 2019/20 and 2025/26. Note that in our simulations, four teams are added to the new design until the 2023/24 season, while four teams are removed from the new design in the 2024/25 and 2025/26 seasons. The fifth-ranked association was always France between the 2019/20 and 2025/26 seasons.

We opted to implement the seeding policy used between the 2003/04 and 2014/15 seasons in the old design, and the official seeding in the new design, which are essentially the same. Both are determined by UEFA club coefficients, except for the automatic assignment of the titleholder to Pot 1. Even though this choice does not allow us to directly compare the old and new designs, it is attractive from an academic perspective for several reasons.
First, the seeding reform is essentially independent of playing in the traditional group stage or in an incomplete round-robin format, and we want to uncover the impact of the latter change.
Second, the effect of the 2015/16 seeding reform has already been studied in the literature \citep{CoronaForrestTenaHorrilloWiper2019, DagaevRudyak2019}.
Third, the seeding regime applied between the 2018/19 and 2023/24 seasons is strongly sensitive to which teams won the national leagues of the highest-ranked associations, as well as the UEFA Europa League. Thus, analysing its effect might require some simplifying assumptions that can be easily debated.

For ease of implementation, the group draw in the old design is simulated by a rejection sampler, which checks the association constraint: a random draw is generated such that each group contains one team from each pot, but it is dismissed if any group contains two teams from the same country. We do not consider the TV pairings as they are not known in the 2024/25 and 2025/26 seasons, where the scheduling of the league phase remains a ``black box''. 
In the new design, the official sequential draw procedure is followed. First, a team from Pot 1 is drawn randomly. Its eight opponents are drawn in home-away pairs sequentially from Pot 1 to Pot 4 with uniform probability. An integer program excludes all pairs that would lead to a deadlock when the remaining teams cannot be drawn without violating a draw constraint (see \citet{DevriesereGoossensSpieksma2026} and the Appendix for the details). The draw continues with choosing another team from Pot 1 randomly. This mechanism is repeated until the last team in Pot 3 is assigned to its opponents from Pot 4.

Naturally, the ranking of the teams is primarily based on their number of points collected in both the old and the new designs.
With respect to tie-breaking rules, the recursive application of head-to-head results in the old format is ignored for the sake of simplicity, and the tie-breaking criteria following the number of goals scored are replaced by a random draw. In the new format, tie-breaking rules that go beyond the match outcomes (i.e., disciplinary record and UEFA club coefficient) are replaced by a random draw.
  
\subsection{Simulating match outcomes} \label{Sec43}

The outcomes of all group and league stage matches are determined by the same approach. We assume that the number of goals scored by a team in a match follows a Poisson distribution \citep{Maher1982, vanEetveldeLey2019}, and the expected number of goals is given by a polynomial of win expectancy, computed from the Elo ratings of the opposing teams. The function is estimated by a least squares regression based on almost eight thousand matches played in UEFA club competitions between 2003/04 and 2023/24, separately for the home and the away team. In particular, the sample contains 2447 UEFA Champions League, 3300 UEFA Europa League (UEFA Cup until the 2008/09 season), and 297 UEFA Europa Conference League (this competition started in 2021/22) games, together with 1898 qualification games of these series. Matches played on a neutral field are excluded.

The win expectancy $W_{ij}$ of team $i$ with Elo $E_i$ playing at home against team $j$ with Elo $E_j$ equals
\begin{equation*} \label{eq1}
W_{ij} = \frac{1}{1 + 10^{-(E_i - E_j)/400}},
\end{equation*}
according to the standard formula of Football Club Elo Ratings (\href{http://clubelo.com/System}{http://clubelo.com/System}). This measure of strength has recently been shown to outperform the official UEFA club coefficient in terms of predictive power \citep{Csato2024c} and is widely used in the literature \citep{BoskerGurtler2024, YildirimBilman2025a, YildirimBilman2025b}.

Let the expected number of goals scored by team $i$ against team $j$ be $\lambda_{ij}^{(f)}$ if the game is played on field $f$ (home: $f = h$; away: $f = a$). Team $i$ scores $k$ goals in this game with the probability of
\begin{equation*} \label{Poisson_dist}
P_{ij}(k) = \frac{ \left( \lambda_{ij}^{(f)} \right)^k \exp \left( -\lambda_{ij}^{(f)} \right)}{k!}.
\end{equation*}
Our estimation for $\lambda_{ij}^{(f)}$ is a cubic polynomial of the win expectancy $W_{ij}$.
For the home team $i$, the expected number of goals equals
\begin{equation*} \label{Exp_goals_home}
\lambda_{ij}^{(h)} = 
2.23998 \cdot W_{ij}^3 - 2.16311 \cdot W_{ij}^2 + 2.48048 \cdot W_{ij} + 0.52717,
\end{equation*}
while for the away team $j$, the expected number of goals equals
\begin{equation*} \label{Exp_goals_away}
\lambda_{ij}^{(a)} = 
-0.79773 \cdot W_{ij}^3 + 2.14427 \cdot W_{ij}^2 -3.06285 \cdot W_{ij} + 2.17402.
\end{equation*}

The idea of approximating the expected number of goals by a polynomial of win expectancy comes from \citet{FootballRankings2020} and has been used in several academic studies \citep{Csato2022a, Csato2023a, Csato2025c, Csato2025d, Stronka2024}. \citet{Gyimesi2024} and \citet{DevriesereGoossensSpieksma2026} have recently followed this approach to evaluate the effect of the 2024/25 UEFA Champions League reform on competitive balance and the competitiveness of matches played in the last rounds of the league phase, respectively.

Crucially, we do not claim that the classical independent Poisson model is the best with respect to predictive performance. Using a more sophisticated simulation model, for example, by introducing correlation or different attack and defense parameters, would probably improve accuracy as demonstrated by \citet{LeyvandeWieleVanEeetvelde2019} and \citet{AlvarezCataldoDuranDuranGalazMonardoSaure2025}. On the other hand, it is highly unlikely that another approach would change our main findings since they are based on comparing two different tournament designs according to the same simulation model.

The old and new Champions League designs can be compared directly by computing the probability of qualifying for the Round of 16, which requires simulating the knockout phase play-offs in the new design, too. Here, the teams aim to win the two-legged clash rather than the individual home and away games. We adopt the solution of \citet{Csato2022b} and \citet{Gyimesi2024}, which is based on the methodology of Football Club Elo Ratings. Hence, team $i$ proceeds to the next round (Round of 16) in the two-legged home-away tie against team $j$ with a probability of 
\begin{equation*}
W_{ij}^\ast = \frac{1}{1 + 10^{-\sqrt{2}(E_i - E_j)/400}}.
\end{equation*}

\begin{table}[t!]
  \centering
  \caption{The strengths of the teams and seedings, 2024/25 season}
  \label{Table2}
\begin{threeparttable}
  \rowcolors{3}{gray!20}{}
    \begin{tabularx}{\textwidth}{ll CCcCc} \toprule \hiderowcolors
    \multirow{2}[0]{*}{Club} & \multirow{2}[0]{*}{Country} & \multirow{2}[0]{*}{Elo} & \multicolumn{2}{c}{Post-2024 seeding} & \multicolumn{2}{c}{Elo-based seeding} \\
          &       &       & Old pot & New pot & Old pot & New pot \\ \bottomrule \showrowcolors
    Real Madrid & Spain & 1987.54 & 1     & 1     & 1     & 1 \\
    Manchester City & England & 2060.21 & 1     & 1     & 1     & 1 \\
    Bayern Munich & Germany & 1908.12 & 1     & 1     & 1     & 1 \\
    Paris Saint-Germain & France & 1895.18 & 1     & 1     & 2     & 1 \\
    Liverpool & England & 1918.22 & 1     & 1     & 1     & 1 \\
    Inter Milan & Italy & 1966.39 & 1     & 1     & 1     & 1 \\
    Borussia Dortmund & Germany & 1870.38 & ---   & 1     & ---   & 2 \\
    RB Leipzig & Germany & 1861.05 & 1     & 1     & 2     & 2 \\
    Barcelona & Spain & 1898.20 & 1     & 1     & 1     & 1 \\
    Bayer Leverkusen & Germany & 1917.84 & 2     & 2     & 1     & 1 \\
    Atl\'etico Madrid & Spain & 1838.46 & 2     & 2     & 2     & 2 \\
    Atalanta & Italy & 1866.29 & 2     & 2     & 2     & 2 \\
    Juventus & Italy & 1833.06 & 2     & 2     & 2     & 2 \\
    Benfica & Portugal & 1759.07 & 2     & 2     & 3     & 3 \\
    Arsenal & England & 1950.36 & 2     & 2     & 1     & 1 \\
    Club Brugge & Belgium & 1708.75 & 2     & 2     & 3     & 3 \\
    Shakhtar Donetsk & Ukraine & 1575.52 & 2     & 2     & 4     & 4 \\
    Milan & Italy & 1817.84 & 3     & 2     & 2     & 2 \\
    Feyenoord & Netherlands & 1748.31 & 3     & 3     & 3     & 3 \\
    Sporting CP & Portugal & 1834.72 & 3     & 3     & 2     & 2 \\
    PSV Eindhoven & Netherlands & 1797.25 & 3     & 3     & 3     & 2 \\
    Dinamo Zagreb & Croatia & 1582.82 & 3     & 3     & 4     & 4 \\
    Red Bull Salzburg & Austria & 1674.05 & 3     & 3     & 4     & 4 \\
    Lille & France & 1770.85 & ---   & 3     & ---   & 3 \\
    Red Star Belgrade & Serbia & 1567.52 & 3     & 3     & 4     & 4 \\
    Young Boys & Switzerland & 1553.15 & 3     & 3     & 4     & 4 \\
    Celtic & Scotland & 1653.25 & 4     & 3     & 4     & 4 \\
    Slovan Bratislava & Slovakia & 1446.40 & ---   & 4     & ---   & 4 \\
    Monaco & France & 1770.45 & 4     & 4     & 3     & 3 \\
    Sparta Prague & Czechia & 1727.75 & 4     & 4     & 3     & 3 \\
    Aston Villa & England & 1777.81 & 4     & 4     & 3     & 3 \\
    Bologna & Italy & 1768.73 & ---   & 4     & ---   & 3 \\
    Girona & Spain & 1800.78 & 4     & 4     & 2     & 2 \\
    VfB Stuttgart & Germany & 1791.27 & 4     & 4     & 3     & 3 \\
    Sturm Graz & Austria & 1606.42 & 4     & 4     & 4     & 4 \\
    Brest & France & 1684.74 & 4     & 4     & 4     & 4 \\ \bottomrule
    \end{tabularx}
\begin{tablenotes} \footnotesize
\item
The teams are ranked according to their UEFA club coefficients, except for the titleholder Real Madrid.
\item
The column Elo shows Football Club Elo Ratings on 2 September 2024 (\url{http://api.clubelo.com/2024-09-02}).
\item
Post-2024 seeding is based on UEFA club coefficients.
The columns Old pot and New pot show the seeding pots of the teams in the old and new designs, respectively (see Section~\ref{Sec413}).
\item
The sign --- indicates teams that are not considered in the old design (see Section~\ref{Sec411}).
\end{tablenotes}
\end{threeparttable}
\end{table}

In contrast to the UEFA club coefficient, the Elo rating of a team is dynamic and changes within a season, as it is updated after each game played by the team. For the baseline 2024/25 season, we use Football Club Elo Ratings on 2 September 2024, which is between the date of the league phase draw (29 August) and the first match (17 September), in the entire season. 
While the outcome of a match may depend on the results of the opposing teams in the previous matches, we think that taking such possible effects into consideration would only obfuscate our results. In fact, the assumption of constant strength during the tournament is in line with the tournament design literature; see, for example, \citet{DevriesereCsatoGoossens2025}. Since the match outcomes are also independent in our model, the qualifying probabilities do not depend on the order of matches; thus, there is no need to generate the schedule of matches, which could be quite challenging due to unknown objectives of the organiser \citep{DevriesereGoossensSpieksma2026}.

The team strength values are reported in Table~\ref{Table2}. The table also gives the assignment of the teams to the pots under both the official and Elo-based seedings.
In order to check the robustness of the numerical results, we use the Football Club Elo Ratings on 1 September of the appropriate year for six further seasons (2019/20--2023/24, 2025/26).

\subsection{The decomposition of draw uncertainty} \label{Sec44}

Following Section~\ref{Sec3}, we generate $K = 1000$ random draws for both the old and new designs separately as described in the Appendix. For each of these draws, $M = 1000$ random sets of match outcomes are simulated as described in Section~\ref{Sec43}.
The effect of the draw is computed as the standard deviation of the draw in the new design ($\sigma_i^n$) minus the standard deviation of the draw in the old design ($\sigma_i^o$), such that both designs use a seeding based on UEFA club coefficients as discussed in Section~\ref{Sec42}.
We label this $\Delta V_i = \sigma_i^n - \sigma_i^o$.

The 2024/25 reform has several elements (see Section~\ref{Sec41}) that may influence draw uncertainty. Therefore, a decomposition procedure is proposed to disentangle these effects for each individual team. For each team $i$, we distinguish three components:

\begin{enumerate}

\item
\emph{Inaccurate seeding effect}: This effect reflects the impact of a seeding system that is based on an inaccurate assessment of team strengths. Hence, we compare the standard deviation of the new and old draws with a seeding based on UEFA club coefficients, with the old and the new draws using a seeding based on Elo ratings. We label this $\Delta V1_i = (\sigma_i^n - \sigma_i^{n, \mathit{Elo}}) - (\sigma_i^o - \sigma_i^{o, \mathit{Elo}})$, where $\sigma_i^{n, \mathit{Elo}}$ ($\sigma_i^{o, \mathit{Elo}}$) is the standard deviation of the draw in the new (old) design with seeding based on Elo ratings. Note that the Elo ratings perfectly reflect team strength due to the simulation model (see Section~\ref{Sec43}). The seeding based on Football Club Elo ratings differs from the seeding based on UEFA club coefficients (Table~\ref{Table2}), although their correlation is strongly positive. 

\item
\emph{Play-off effect}: This effect indicates the impact of introducing the knockout phase play-offs after the league phase, instead of letting the best 16 teams of the league phase qualify directly for the Round of 16. We denote it as $\Delta V2_i = \sigma_i^{n, \mathit{Elo}} - \sigma_i^{n, \mathit{Elo}, T16}$, where $\sigma_i^{n, \mathit{Elo}, T16}$ is the standard deviation of the draw in the new design, assuming that the top 16 teams in the league phase qualify directly for the Round of 16 without the knockout phase play-offs. This comparison is done using the (accurate) seeding based on Elo ratings.

\item
\emph{First stage effect}: This effect reflects the impact of changing from a group stage to a league phase, where each team plays two additional matches against two different teams from its seeding pot, and they are ranked in a single league, with the best 16 qualifying directly for the Round of 16. We denote this effect as $\Delta V3_i = \sigma_i^{n, \mathit{Elo}, T16} - \sigma_i^{o, \mathit{Elo}}$, again using the seeding based on Elo ratings. 
\end{enumerate}
Obviously, $\Delta V_i = \Delta V1_i + \Delta V2_i + \Delta V3_i$, hence, the effect of the 2024/25 reform is indeed decomposed into three elements.

The contribution of inaccurate seeding is removed first since we want to compare the incomplete round-robin format to the group format in its pure form. Then, we investigate the impact of the additional play-offs (assuming that team strengths are known and the tournament is seeded perfectly), and finally, the first stage effect, which is the impact of changing the group stage to an incomplete league. There can be other (valid) ways to decompose $\Delta V_i$, for example, by introducing new components or isolating the effects in a different order. We also remind the reader that the modified definition of Pot 1 is \emph{not} studied, as this has already been addressed by \citet{CoronaForrestTenaHorrilloWiper2019} and \citet{DagaevRudyak2019}.

While the magnitude of the three effects is influenced by the standard deviation of the draw using Elo-based seedings, the total effect of the reform $\Delta V_i$ is, naturally, independent of both $\sigma_i^{n, \mathit{Elo}}$ and $\sigma_i^{o, \mathit{Elo}}$. Furthermore, by considering a seeding based on Elo ratings, we do not claim that the Elo ratings reflect the actual strength of the teams. However, if the team strengths used for simulating match outcomes are fully aligned with the seeding, then the seeding can be called ``perfect’’ as any bias due to inaccurate seeding is removed.

Naturally, the number of simulated draws affects the accuracy of the standard deviations computed. Therefore, we generate bootstrap confidence intervals as follows. The 1000 random draws give 1000 qualifying probabilities for each team. For each team, one thousand samples of 1000 are generated with replacement from its 1000 qualifying probabilities, and the standard deviation is computed from each sample. The 2.5th and 97.5th percentiles provide the lower and upper bounds of the 95\% confidence interval for the standard deviation of the qualifying probability.

Note that our computation of the standard deviation contains an inherent bias because the qualifying probabilities are estimated from $M = 1000$ simulations. Consequently, even if the draw did not have any effect on the qualifying probabilities, we find a non-zero standard deviation. Indeed, according to numerical experiments, increasing the value of $M$ reduces the standard deviation.
However, as we have discussed in Section~\ref{Sec3}, the interpretation of $\sigma_i$ on its own is challenging, and the choice of $M$ does not influence our findings on the order of different tournament designs with respect to the implied standard deviations.

\section{Results} \label{Sec5}

The discussion of our findings is divided into three parts. Section~\ref{Sec51} compares draw uncertainty in the old and new designs of the UEFA Champions League based on the 2024/25 season. Section~\ref{Sec52} applies the decomposition method proposed in Section~\ref{Sec44} to disentangle the first stage, play-off, and seeding effects in this season. Finally, Section~\ref{Sec53} provides a sensitivity analysis by investigating six additional seasons and the impact of the number of random draws. 

\subsection{The overall effect of the 2024/25 reform} \label{Sec51}

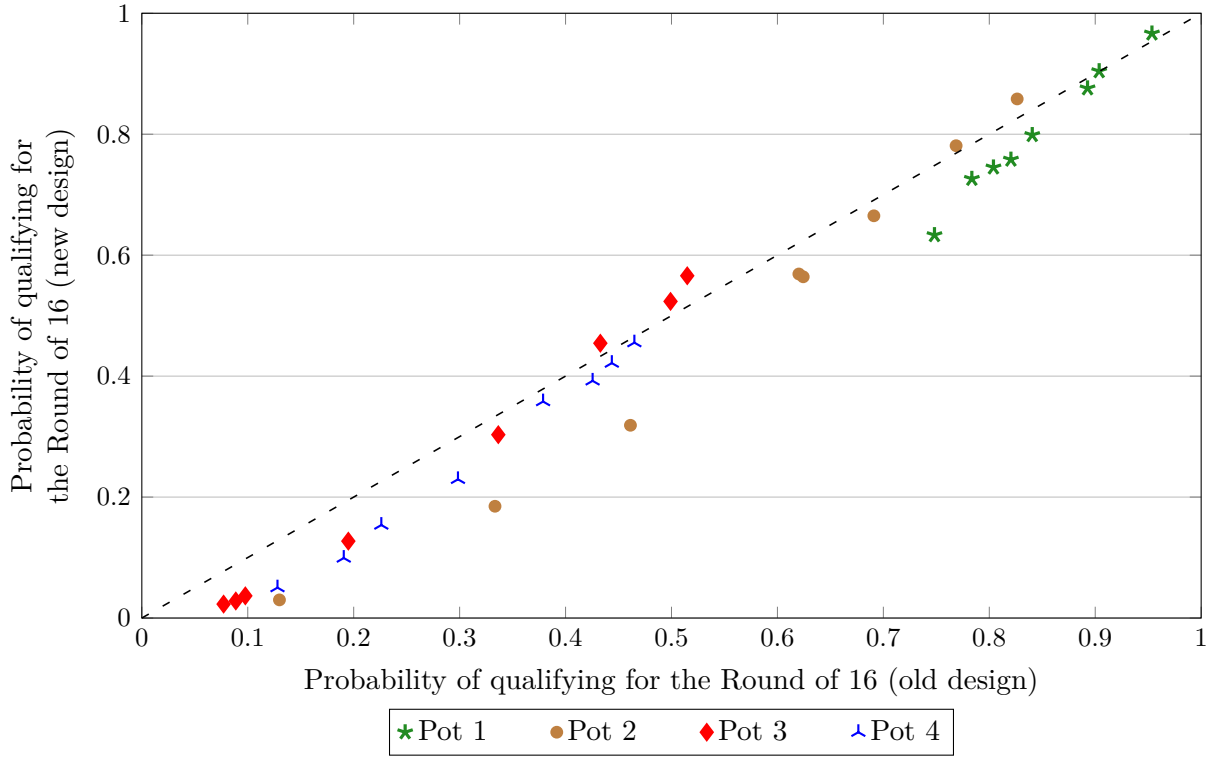
\begin{figure}[t]
\centering

\begin{tikzpicture}
\begin{axis}[width = 0.97\textwidth, 
height = 0.6\textwidth,
xmin = 0,
xmax = 1,
ymin = 0,
ymax = 1,
ymajorgrids,
scaled ticks = false,
tick label style = {/pgf/number format/fixed},
xlabel = Probability of qualifying for the Round of 16 (old design),
xlabel style = {font = \small},
ylabel = {Probability of qualifying for \\ the Round of 16 (new design)},
ylabel style = {align = center, font = \small}, 
legend entries = {Pot 1$\qquad$,Pot 2$\qquad$,Pot 3$\qquad$,Pot 4},
legend style = {at = {(0.5,-0.15)},anchor = north,legend columns = 4,font = \small}
]
\addplot[black,thick,only marks,mark=otimes*, mark size=2pt] coordinates {
(0.803936,0.745155)
(0.820415,0.758079)
(0.892738,0.875939)
(0.840561,0.79874)
(0.953543,0.966646)
(0.783527,0.72632)
(0.748254,0.633263)
(0.903729,0.904331)
};

\addplot[ForestGreen,very thick,only marks,mark=x, mark size=3pt] coordinates {
(0.826322,0.858347)
(0.691045,0.665117)
(0.620255,0.568907)
(0.76868,0.781026)
(0.461255,0.318732)
(0.333389,0.184822)
(0.624344,0.564232)
(0.129915,0.030017)
};

\addplot[red,thick,only marks,mark=diamond*, mark size=3pt] coordinates {
(0.097706,0.036689)
(0.336516,0.303168)
(0.499111,0.523559)
(0.432838,0.454411)
(0.194942,0.127143)
(0.088715,0.028195)
(0.514871,0.565955)
(0.0772,0.023032)
};

\addplot[blue,thick,only marks,mark=star, mark size=3pt] coordinates {
(0.425545,0.392184)
(0.226081,0.153757)
(0.190636,0.099204)
(0.464961,0.455366)
(0.378787,0.357942)
(0.298436,0.229349)
(0.12806,0.050256)
(0.443687,0.421509)
};
\end{axis}

\draw [black,loosely dashed,semithick] (rel axis cs:0,0) -- (rel axis cs:1,1);
\end{tikzpicture}

\caption{The probability of qualification for the Round of 16  in the old and new \\ designs of the UEFA Champions League (pots according to the old design)}
\label{Fig1}

\end{figure}


Figure~\ref{Fig1} compares the probability of reaching the Round of 16 for the old and new designs of the UEFA Champions League for the 2024/25 season. Unsurprisingly, the new design tends to reduce the chances of the teams due to the increase in the number of participants from 32 to 36. Note that the additional four teams are not underdogs except for Slovan Bratislava; Borussia Dortmund has the 10th-highest Elo rating (Table~\ref{Table2}). In absolute terms, the greatest loser is Club Brugge, its chance of qualification for the Round of 16 has decreased from 34.3\% to 18.4\%. In relative terms, the greatest loser is Shakhtar Donetsk, its probability of reaching the Round of 16 is less than one-fourth in the new design compared to the old design. These teams are the weakest in Pot 2, which could have benefited from the higher variance in the strength of their opponents in the former groups.

Two sets of teams gain in the new design:
(a) the strongest teams because the higher number of matches played is favourable for them \citep{LasekGagolewski2018, SziklaiBiroCsato2022};
(b) the strongest teams from Pots 2 and 3 that can exploit the two additional matches against weak teams in their own pots, as well as the more diverse set of opponents in the league phase.
In the 2024/25 season, the champions of some lower-ranked associations (Dinamo Zagreb, Red Bull Salzburg, Red Star Belgrade, Shakhtar Donetsk) have a higher UEFA club coefficient than several relatively strong teams from higher-ranked associations (Aston Villa, Girona, Monaco, VfB Stuttgart) because these champions have been regular participants in UEFA club competitions in previous years. Consequently, the average strength of Pot 3 according to the Elo ratings (1686.88) is smaller than the average strength of Pot 4 (1708.26).

Note that changes in qualification probabilities also have non-marginal financial implications. Advancing to the Round of 16 yields an additional 11 million euros in prize money, while performance-based rewards during the league phase depend on the number of wins and draws, which in turn are influenced by the outcome of the league phase draw \citep{UEFA2025a}. Moreover, 35\% of the net revenue are allocated by the TV market pool of the associations of the clubs and their coefficient ranking over multiple seasons. Consequently, the financial rewards are not solely a function of sporting performance, but depend on the identities of the participating teams, too. This makes a comprehensive assessment of the financial implications of the Champions League draw procedure considerably complex, although it remains a promising avenue for future research.

\begin{figure}[t]
\centering

\begin{tikzpicture}
\begin{axis}[width = 0.97\textwidth, 
height = 0.6\textwidth,
xmin = 0,
xmax = 0.155,
ymin = 0,
ymax = 0.155,
ymajorgrids,
scaled ticks = false,
tick label style = {/pgf/number format/fixed},
xlabel = {Standard deviation of reaching the Round of 16 (old design, $\sigma_i^o$)},
xlabel style = {font = \small},
ylabel = {Standard deviation of reaching the \\ Round of 16 (new design, $\sigma_i^n$)},
ylabel style = {align = center, font = \small}, 
legend entries = {Pot 1$\qquad$,Pot 2$\qquad$,Pot 3$\qquad$,Pot 4},
legend style = {at = {(0.5,-0.15)},anchor = north,legend columns = 4,font = \small}
]
\addplot[black,thick,only marks,mark=otimes*, mark size=2pt] coordinates {
(0.0856737526661223,0.0502465091990149)
(0.0720836932588351,0.0478992598909221)
(0.0516401981606003,0.0303407620346102)
(0.0673972749573132,0.0403000309242126)
(0.0202774092113446,0.00928735371023087)
(0.0855442873375896,0.0540657627959387)
(0.0955241348788061,0.0584048367513852)
(0.0449403960618751,0.024314988333488)
};

\addplot[ForestGreen,very thick,only marks,mark=x, mark size=3pt] coordinates {
(0.0542946058114452,0.0295049340302962)
(0.0935502251179526,0.0630040857867633)
(0.102363197484248,0.0618819822786143)
(0.0724558033800154,0.0424219158330095)
(0.11500793363851,0.0584310907682577)
(0.117003801513321,0.0475814492245698)
(0.106056643713045,0.0669668928705875)
(0.0672746434598702,0.0122867275820161)
};

\addplot[red,thick,only marks,mark=diamond*, mark size=3pt] coordinates {
(0.0532740996516477,0.0148419190659955)
(0.111212094098951,0.0593388310710017)
(0.126894353128324,0.0719488231464014)
(0.116508411567431,0.0634194300124426)
(0.0850467258722026,0.0387321924725897)
(0.0500089659128408,0.0117890609022499)
(0.115878963980928,0.0605751956606772)
(0.0454536911376627,0.0104810700230858)
};

\addplot[blue,thick,only marks,mark=star, mark size=3pt] coordinates {
(0.129506563250381,0.0607982612627415)
(0.109363571767887,0.0432096635722384)
(0.107745638550368,0.0317833693394292)
(0.139419601578488,0.0637216642476834)
(0.130838045786481,0.0632469035353258)
(0.123062879766306,0.0526139660967355)
(0.084113420257813,0.020122111304815)
(0.140860195320132,0.0655045553054577)
};
\end{axis}

\draw [black,loosely dashed,semithick] (rel axis cs:0,0) -- (rel axis cs:1,1);
\end{tikzpicture}

\caption{The standard deviation of reaching the Round of 16 in the old and \\ new designs of the UEFA Champions League (pots according to the old design)}
\label{Fig2}

\end{figure}
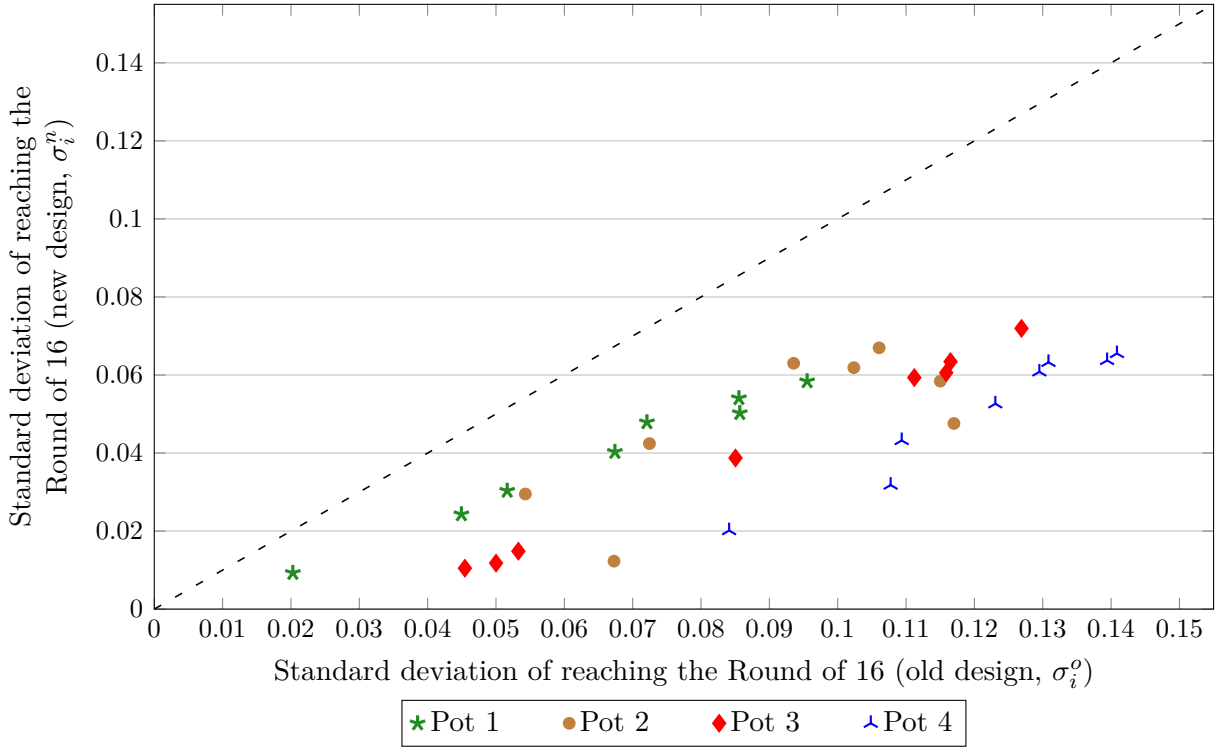


Figure~\ref{Fig2} quantifies draw uncertainty by the standard deviation of qualifying probabilities in the old and new designs in the 2024/25 season. The impact of the draw has decreased for all teams. As expected, the reduction in absolute (but not in relative) terms is the smallest for the strongest and the weakest teams, as these teams have either a high or a low chance of reaching the Round of 16 under any draw. However, for the middle teams, the effect of the draw is substantially smaller in the new design compared to the old design. The most striking example is again Shakhtar Donetsk, with the fourth-lowest Elo rating overall, but assigned to Pot 2: its standard deviation is reduced by more than 80\% as it is less likely in the new design that Donetsk could play against only weak teams from Pots 3 and 4.

\begin{table}[t!]
  \centering
  \caption{The qualifying probabilities and their standard deviations \\ for each team, 2024/25 season}
  \label{Table3}
\centerline{
\begin{threeparttable}
  \rowcolors{3}{gray!20}{}
    \begin{tabularx}{1.08\textwidth}{l Ccc ccc} \toprule \hiderowcolors
    \multirow{2}[0]{*}{Club} & \multirow{2}[0]{*}{Elo} & \multicolumn{2}{c}{Qualifying for R16 (\%)} & \multicolumn{3}{c}{Standard deviation} \\
          &       & Old design  & New design  & Old ($\sigma_i^o$)   & New ($\sigma_i^n$)  & Change (\%) \\ \bottomrule \showrowcolors
    Manchester City & 2060  & 95.35 & 96.66 & 0.0203 & 0.0093 & $-$54.20 \\
    Real Madrid & 1988  & 90.37 & 90.43 & 0.0449 & 0.0243 & $-$45.90 \\
    Inter Milan & 1966  & 89.27 & 87.59 & 0.0516 & 0.0303 & $-$41.25 \\
    Arsenal & 1950  & 82.63 & 85.83 & 0.0543 & 0.0295 & $-$45.66 \\
    Liverpool & 1918  & 84.06 & 79.87 & 0.0674 & 0.0403 & $-$40.21 \\
    Bayer Leverkusen & 1918  & 76.87 & 78.10 & 0.0725 & 0.0424 & $-$41.45 \\
    Bayern Munich & 1908  & 82.04 & 75.81 & 0.0721 & 0.0479 & $-$33.55 \\
    Barcelona & 1898  & 80.39 & 74.52 & 0.0857 & 0.0502 & $-$41.35 \\
    Paris Saint-Germain & 1895  & 78.35 & 72.63 & 0.0855 & 0.0541 & $-$36.80 \\
    Borussia Dortmund & 1870  & ---   & 66.24 & ---   & 0.0562 & --- \\
    Atalanta & 1866  & 69.10 & 66.51 & 0.0936 & 0.0630 & $-$32.65 \\
    RB Leipzig & 1861  & 74.83 & 63.33 & 0.0955 & 0.0584 & $-$38.86 \\
    Atl\'etico Madrid & 1838  & 62.03 & 56.89 & 0.1024 & 0.0619 & $-$39.55 \\
    Sporting CP & 1835  & 51.49 & 56.60 & 0.1159 & 0.0606 & $-$47.73 \\
    Juventus & 1833  & 62.43 & 56.42 & 0.1061 & 0.0670 & $-$36.86 \\
    Milan & 1818  & 49.91 & 52.36 & 0.1269 & 0.0719 & $-$43.30 \\
    Girona & 1801  & 46.50 & 45.54 & 0.1394 & 0.0637 & $-$54.30 \\
    PSV Eindhoven & 1797  & 43.28 & 45.44 & 0.1165 & 0.0634 & $-$45.57 \\
    VfB Stuttgart & 1791  & 44.37 & 42.15 & 0.1409 & 0.0655 & $-$53.50 \\
    Aston Villa & 1778  & 42.55 & 39.22 & 0.1295 & 0.0608 & $-$53.05 \\
    Lille & 1771  & ---   & 36.38 & ---   & 0.0652 & --- \\
    Monaco & 1770  & 37.88 & 35.79 & 0.1308 & 0.0632 & $-$51.66 \\
    Bologna & 1769  & ---   & 36.85 & ---   & 0.0665 & --- \\
    Benfica & 1759  & 46.13 & 31.87 & 0.1150 & 0.0584 & $-$49.19 \\
    Feyenoord & 1748  & 33.65 & 30.32 & 0.1112 & 0.0593 & $-$46.64 \\
    Sparta Prague & 1728  & 29.84 & 22.93 & 0.1231 & 0.0526 & $-$57.25 \\
    Club Brugge & 1709  & 33.34 & 18.48 & 0.1170 & 0.0476 & $-$59.33 \\
    Brest & 1685  & 22.61 & 15.38 & 0.1094 & 0.0432 & $-$60.49 \\
    Red Bull Salzburg & 1674  & 19.49 & 12.71 & 0.0850 & 0.0387 & $-$54.46 \\
    Celtic & 1653  & 19.06 & 9.92  & 0.1077 & 0.0318 & $-$70.50 \\
    Sturm Graz & 1606  & 12.81 & 5.03  & 0.0841 & 0.0201 & $-$76.08 \\
    Dinamo Zagreb & 1583  & 9.77  & 3.67  & 0.0533 & 0.0148 & $-$72.14 \\
    Shakhtar Donetsk & 1576  & 12.99 & 3.00  & 0.0673 & 0.0123 & $-$81.74 \\
    Red Star Belgrade & 1568  & 8.87  & 2.82  & 0.0500 & 0.0118 & $-$76.43 \\
    Young Boys & 1553  & 7.72  & 2.30  & 0.0455 & 0.0105 & $-$76.94 \\
    Slovan Bratislava & 1446  & ---   & 0.39  & ---   & 0.0024 & --- \\ \bottomrule
    \end{tabularx}
\begin{tablenotes} \footnotesize
\item
The teams are ranked according to Football Club Elo Ratings on 2 September 2024 (\url{http://api.clubelo.com/2024-09-02}), reported in column Elo.
\item
The columns Old (design) and New (design) show the simulation results for the old and new designs, respectively.
\item
The last column Change shows the relative change of the standard deviation of reaching the Round of 16 for each team due to the 2024/25 reform (except for the modification in the seeding system, see Section~\ref{Sec42}).
\item
The sign --- indicates teams that are not considered in the old design (see Section~\ref{Sec411}).
\end{tablenotes}
\end{threeparttable}
}
\end{table}

Table~\ref{Table3} reports the probability of reaching the Round of 16 and its standard deviation in the old and new UEFA Champions League for each team in the 2024/25 season. The latter is reduced by at least 30\% but sometimes by more than 75\%, and the new design decreases the uncertainty of the draw by 53\% on average.

\subsection{The decomposition of the change in draw uncertainty} \label{Sec52}

Since Figure~\ref{Fig2} is influenced by several modifications in the 2024/25 season, it obscures the effects of individual elements on the overall change in draw uncertainty. This has motivated the decomposition method presented in Section~\ref{Sec44}.

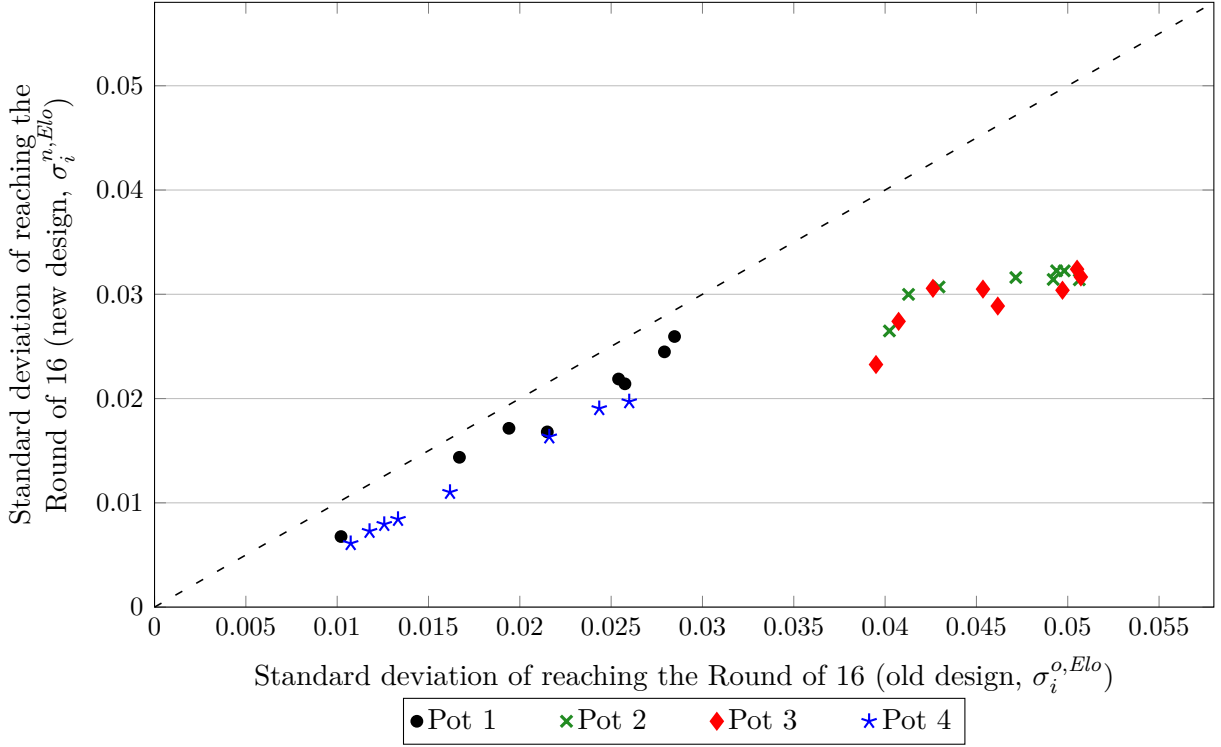
\begin{figure}[t]
\centering

\begin{tikzpicture}
\begin{axis}[width = 0.97\textwidth, 
height = 0.6\textwidth,
xmin = 0,
xmax = 0.058,
ymin = 0,
ymax = 0.058,
ymajorgrids,
scaled ticks = false,
tick label style = {/pgf/number format/.cd, fixed, precision=3},
xlabel = {Standard deviation of reaching the Round of 16 (old design, $\sigma_i^{o, \mathit{Elo}}$)},
xlabel style = {font = \small},
ylabel = {Standard deviation of reaching the \\ Round of 16 (new design, $\sigma_i^{n, \mathit{Elo}}$)},
ylabel style = {align = center, font = \small}, 
legend entries = {Pot 1$\qquad$,Pot 2$\qquad$,Pot 3$\qquad$,Pot 4},
legend style = {at = {(0.5,-0.15)},anchor = north,legend columns = 4,font = \small}
]
\addplot[black,thick,only marks,mark=otimes*, mark size=2pt] coordinates {
(0.0215044086315014,0.016803251866031)
(0.0284710512056853,0.0259508499015083)
(0.0254074499608582,0.021876755322109)
(0.0279177335011968,0.0244802610930354)
(0.0194047804549422,0.0171469525567076)
(0.0257538159120942,0.0214140963508275)
(0.0102105752456386,0.00676614067690436)
(0.0166930451912693,0.0143657218741338)
};

\addplot[ForestGreen,very thick,only marks,mark=x, mark size=3pt] coordinates {
(0.0429542303493487,0.0307032321338237)
(0.0471571723378629,0.0316093231232228)
(0.0498093422017453,0.0322512167985176)
(0.0492029812548423,0.0314239250104608)
(0.0493892507950272,0.0322578532003153)
(0.0402367905692703,0.0264858069344502)
(0.0412819236618307,0.0299907543310492)
(0.0506334009552222,0.0313879804434168)
};

\addplot[red,thick,only marks,mark=diamond*, mark size=3pt] coordinates {
(0.0453582280060393,0.0304967631653519)
(0.0497118454553699,0.0303882156048631)
(0.0395013604382998,0.0232643558744757)
(0.046169428110949,0.0288736973974827)
(0.0426191784220364,0.0305704129698823)
(0.050511622424943,0.0324065552600677)
(0.0407389880722282,0.0274028460560572)
(0.0507193830287279,0.0316624583737629)
};

\addplot[blue,thick,only marks,mark=star, mark size=3pt] coordinates {
(0.0259899033294677,0.0197062759625535)
(0.0216139586195405,0.016333020497096)
(0.0133307263217075,0.00842453992054598)
(0.0243479096578064,0.019047769647088)
(0.0117713610062703,0.00727562270088861)
(0.0125771564299671,0.00794067694534605)
(0.0161772461900749,0.0110159073259634)
(0.0107398473136409,0.00610496638931081)
};
\end{axis}

\draw [black,loosely dashed,semithick] (rel axis cs:0,0) -- (rel axis cs:1,1);
\end{tikzpicture}

\caption{The standard deviation of reaching the Round of 16 \\ in the old and new designs of the UEFA Champions League, \\ seeding based on Elo ratings (pots according to the old design)}
\label{Fig3}

\end{figure}


Figure~\ref{Fig3} attempts to filter out the effect of the inaccurate seeding by assuming that the pots in the 2024/25 season are formed according to the actual strengths of the teams (which is given by their Elo ratings in our simulation), in both the old and new designs. Unsurprisingly, the magnitude of standard deviations is substantially reduced compared to Figure~\ref{Fig2}. The decrease due to the new format is also smaller for all teams, and the standard deviation of the draw is essentially unchanged for the eight strongest teams. The probable reason is that they face two new opponents of comparable strength, and two losses in these matches -- which have a non-marginal probability -- can really harm them.

\begin{figure}[t]
\centering

\begin{tikzpicture}
\begin{axis}[width = 0.97\textwidth, 
height = 0.6\textwidth,
xmin = 0,
xmax = 0.058,
ymin = 0,
ymax = 0.058,
ymajorgrids,
scaled ticks = false,
tick label style = {/pgf/number format/.cd, fixed, precision=3},
xlabel = {Standard deviation of reaching the Round of 16 (old design, $\sigma_i^{o, \mathit{Elo}}$)},
xlabel style = {font = \small},
ylabel = {Standard deviation of reaching the Round of 16 \\ (new design without play-offs, $\sigma_i^{n, \mathit{Elo}, T16}$)},
ylabel style = {align = center, font = \small}, 
legend entries = {Pot 1$\qquad$,Pot 2$\qquad$,Pot 3$\qquad$,Pot 4},
legend style = {at = {(0.5,-0.15)},anchor = north,legend columns = 4,font = \small}
]
\addplot[black,thick,only marks,mark=otimes*, mark size=2pt] coordinates {
(0.0215044086315014,0.0225104996300303)
(0.0284710512056853,0.0348882166161966)
(0.0254074499608582,0.0300536040953347)
(0.0279177335011968,0.0330631711165805)
(0.0194047804549422,0.0238438814307084)
(0.0257538159120942,0.0279144871633637)
(0.0102105752456386,0.0100481585123588)
(0.0166930451912693,0.0209888485335133)
};

\addplot[ForestGreen,very thick,only marks,mark=x, mark size=3pt] coordinates {
(0.0429542303493487,0.039458218345422)
(0.0471571723378629,0.0408471801122565)
(0.0498093422017453,0.0414541834478177)
(0.0492029812548423,0.0407641572035104)
(0.0493892507950272,0.0415807131577455)
(0.0402367905692703,0.0349589182464928)
(0.0412819236618307,0.0388068257620451)
(0.0506334009552222,0.0395762923846956)
};

\addplot[red,thick,only marks,mark=diamond*, mark size=3pt] coordinates {
(0.0453582280060393,0.0392293897610072)
(0.0497118454553699,0.0394067330491835)
(0.0395013604382998,0.0315155102510873)
(0.046169428110949,0.0375932208550193)
(0.0426191784220364,0.0389498072843434)
(0.050511622424943,0.0408646159389249)
(0.0407389880722282,0.0361314283708258)
(0.0507193830287279,0.0410594304981551)
};

\addplot[blue,thick,only marks,mark=star, mark size=3pt] coordinates {
(0.0259899033294677,0.0271277376526343)
(0.0216139586195405,0.0231482369063363)
(0.0133307263217075,0.0132618863148735)
(0.0243479096578064,0.0267083814867476)
(0.0117713610062703,0.0115945974785971)
(0.0125771564299671,0.0125971609286959)
(0.0161772461900749,0.016721090690319)
(0.0107398473136409,0.010224049451523)
};
\end{axis}

\draw [black,loosely dashed,semithick] (rel axis cs:0,0) -- (rel axis cs:1,1);
\end{tikzpicture}

\caption{The standard deviation of finishing in the Top 16 in the \\ first stage of the old and new UEFA Champions League designs, \\ seeding based on Elo ratings (pots according to the old design)}
\label{Fig4}

\end{figure}
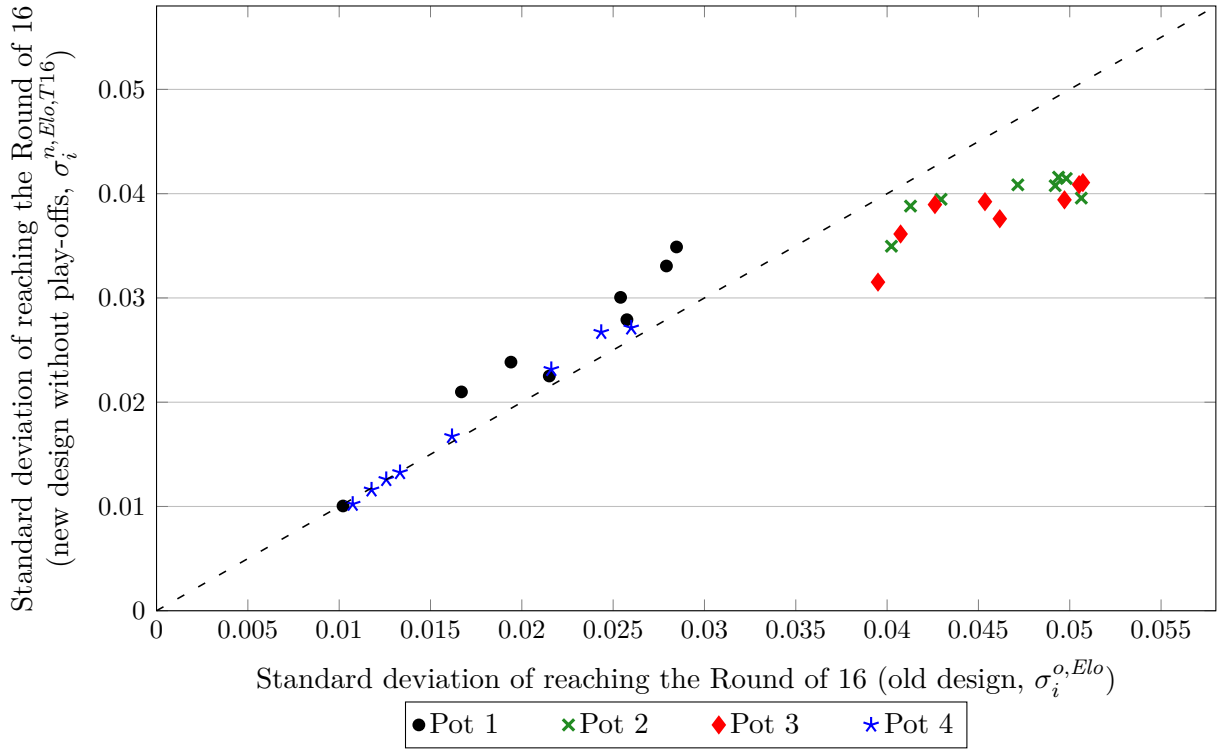


Nonetheless, Figure~\ref{Fig3} still contains the effect of the novel knockout phase play-offs, which is not an essential part of the new design: UEFA could decide that the teams ranked 1--16 in the league phase qualify directly for the Round of 16 in order to reduce the number of matches.
Hence, Figure~\ref{Fig4} removes the play-offs and retains the perfect seeding to compare the effect of the draw in the group stage and the incomplete round-robin league phase. The variance of reaching the Round of 16 decreases somewhat only for the middle teams (those in Pots 2 and 3 in the Elo-based seeding), but it becomes higher for the weakest (Pot 4) and, especially, for the strongest teams (Pot 1). Since the middle teams qualify for the Round of 16 with a probability closer to 0.5, they face a higher standard deviation due to the variance of the binomial distribution, which can be reduced by playing against a more diverse set of opponents in an incomplete round-robin tournament. On the other hand, all teams play two additional matches against their peers in their pots, yielding a new source of uncertainty that becomes dominant for the strongest and the weakest teams.

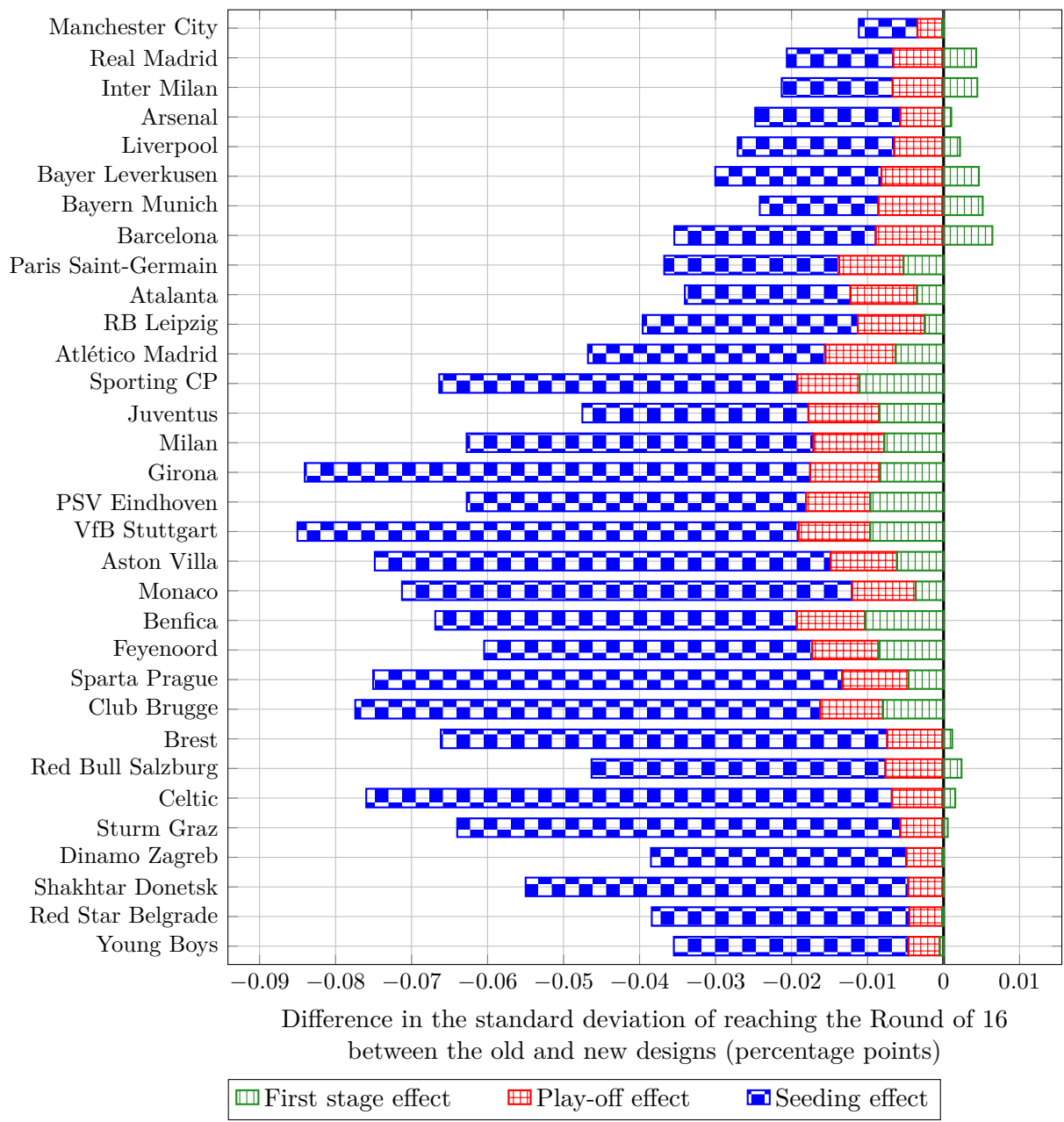
\begin{figure}[t!]
\centering

\begin{tikzpicture}
\begin{axis}[
width = 0.88\textwidth, 
height = \textwidth, 
xmajorgrids,
ymajorgrids,
xbar stacked,
bar width = 8pt,
scaled x ticks = false,
xlabel = {Difference in the standard deviation of reaching the Round of 16 \\ between the old and new designs (percentage points)},
xlabel style = {align=center, font=\small},
xticklabel style = {/pgf/number format/fixed,/pgf/number format/precision=5},
extra x ticks = 0,
extra x tick labels = ,
extra x tick style = {grid = major, major grid style = {black,very thick}},
symbolic y coords = {Manchester City,Real Madrid,Inter Milan,Arsenal,Liverpool,Bayer Leverkusen,Bayern Munich,Barcelona,Paris Saint-Germain,Atalanta,RB Leipzig,Atl\'etico Madrid,Sporting CP,Juventus,Milan,Girona,PSV Eindhoven,VfB Stuttgart,Aston Villa,Monaco,Benfica,Feyenoord,Sparta Prague,Club Brugge,Brest,Red Bull Salzburg,Celtic,Sturm Graz,Dinamo Zagreb,Shakhtar Donetsk,Red Star Belgrade,Young Boys},
ytick = data,
y dir = reverse,
enlarge y limits = 0.02,
legend style = {font=\small,at={(0,-0.12)},anchor=north west,legend columns=3},
legend entries = {First stage effect$\qquad$, Play-off effect$\qquad$, Seeding effect},
]
\addplot [ForestGreen, thick, pattern = vertical lines, pattern color = ForestGreen] coordinates{
(-0.000162416733279731,Manchester City)
(0.00429580334224404,Real Madrid)
(0.00443910097576626,Inter Milan)
(0.00100609099852888,Arsenal)
(0.00216067125126949,Liverpool)
(0.00464615413447651,Bayer Leverkusen)
(0.00514543761538378,Bayern Munich)
(0.0064171654105113,Barcelona)
(-0.0052778723227775,Paris Saint-Germain)
(-0.00349601200392668,Atalanta)
(-0.00247509789978565,RB Leipzig)
(-0.00630999222560642,Atl\'etico Madrid)
(-0.0110571085705265,Sporting CP)
(-0.00843882405133189,Juventus)
(-0.00780853763728172,Milan)
(-0.00835515875392762,Girona)
(-0.00964700648601805,PSV Eindhoven)
(-0.00965995253057279,VfB Stuttgart)
(-0.00612883824503205,Aston Villa)
(-0.003669371137693,Monaco)
(-0.0103051124061864,Benfica)
(-0.00857620725592979,Feyenoord)
(-0.00460755970140236,Sparta Prague)
(-0.00798585018721253,Club Brugge)
(0.00113783432316657,Brest)
(0.00236047182894114,Red Bull Salzburg)
(0.00153427828679588,Celtic)
(0.000543844500244087,Sturm Graz)
(-6.88400068339801E-05,Dinamo Zagreb)
(2.00044987287984E-05,Shakhtar Donetsk)
(-0.000176763527673139,Red Star Belgrade)
(-0.000515797862117959,Young Boys)
};
\addplot [red, thick, pattern = grid, pattern color = red] coordinates{
(-0.00328201783545446,Manchester City)
(-0.00662312665937954,Real Madrid)
(-0.00669692887400079,Inter Milan)
(-0.00570724776399929,Arsenal)
(-0.00650039081253619,Liverpool)
(-0.00817684877322575,Bayer Leverkusen)
(-0.00858291002354511,Bayern Munich)
(-0.00893736671468835,Barcelona)
(-0.00847311131204257,Paris Saint-Germain)
(-0.00875498621159828,Atalanta)
(-0.00881607143099589,RB Leipzig)
(-0.0092378569890337,Atl\'etico Madrid)
(-0.00818831194127885,Sporting CP)
(-0.00934023219304966,Juventus)
(-0.00932285995743014,Milan)
(-0.00920296664930011,Girona)
(-0.00845806067885718,PSV Eindhoven)
(-0.00939697212439222,VfB Stuttgart)
(-0.00873262659565536,Aston Villa)
(-0.00837939431446104,Monaco)
(-0.00901851744432034,Benfica)
(-0.00871952345753652,Feyenoord)
(-0.00872858231476857,Sparta Prague)
(-0.00825115437661152,Club Brugge)
(-0.00742146169008079,Brest)
(-0.00766061183965955,Red Bull Salzburg)
(-0.00681521640924033,Celtic)
(-0.00570518336435562,Sturm Graz)
(-0.00483734639432751,Dinamo Zagreb)
(-0.00465648398334989,Shakhtar Donetsk)
(-0.00431897477770851,Red Star Belgrade)
(-0.00411908306221217,Young Boys)
};
\addplot [blue, thick, pattern = crosshatch dots, pattern color = blue] coordinates{
(-0.00770803766565929,Manchester City)
(-0.0140022810690075,Real Madrid)
(-0.0146025072519893,Inter Milan)
(-0.0190824240171497,Arsenal)
(-0.0205968532205644,Liverpool)
(-0.0218570387737801,Bayer Leverkusen)
(-0.0156015233443679,Bayern Munich)
(-0.0264898767524191,Barcelona)
(-0.0230054132296084,Paris Saint-Germain)
(-0.0217911531195911,Atalanta)
(-0.028303226696425,RB Leipzig)
(-0.0312433582166002,Atl\'etico Madrid)
(-0.0471154563789723,Sporting CP)
(-0.0297495186494075,Juventus)
(-0.0456226700244927,Milan)
(-0.0664949706815044,Girona)
(-0.0446309208761315,PSV Eindhoven)
(-0.0659586678902817,VfB Stuttgart)
(-0.0599756753919842,Aston Villa)
(-0.0592117479366939,Monaco)
(-0.0475583254259315,Benfica)
(-0.0431537395704125,Feyenoord)
(-0.0617203313548024,Sparta Prague)
(-0.0611711979121395,Club Brugge)
(-0.0587324465055681,Brest)
(-0.0386539215599534,Red Bull Salzburg)
(-0.0691470528016989,Celtic)
(-0.0582861255886424,Sturm Graz)
(-0.0335948341913247,Dinamo Zagreb)
(-0.0503314318945042,Shakhtar Donetsk)
(-0.0339009302328824,Red Star Belgrade)
(-0.0308535380523647,Young Boys)
};
\end{axis}
\end{tikzpicture}

\caption{Decomposition of changes in the \\ draw uncertainty of the UEFA Champions League \\ \vspace{0.2cm}
\footnotesize{\emph{Note:} The teams are ranked according to their Elo rating.}
}
\label{Fig5}

\end{figure}


Figure~\ref{Fig5} decomposes the changes in draw uncertainty into three components as described in Section~\ref{Sec44}. The first stage effect, shown in Figure~\ref{Fig4}, is positive for the eight bottom teams and, especially, for the eight strongest teams (except for the outstanding Manchester City). The play-off effect is somewhat more important, more homogeneous, and always negative; that is, the introduction of the knockout stage play-offs has removed a substantial amount of uncertainty in the new design. Furthermore, the seeding effect dominates and largely drives the reduced standard deviation seen in Figure~\ref{Fig2}. The effect of inaccurate seeding is generally higher for middle teams, especially if they are assigned to a different pot than implied by their Elo rating. For example, Juventus and Sporting CP are equally strong, but the seeding effect is almost doubled for Sporting CP since it is assigned to Pot 3 rather than Pot 2 according to UEFA club coefficients.

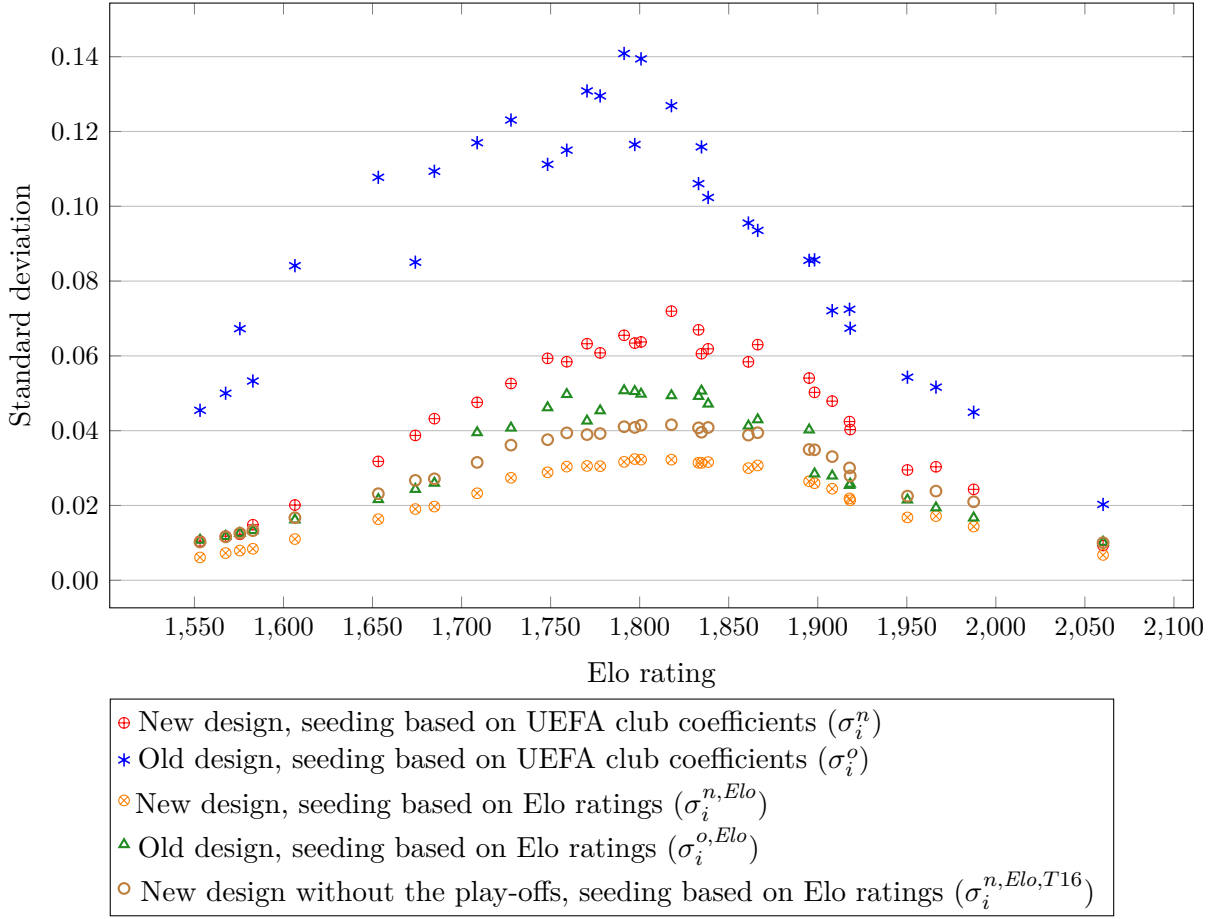
\begin{figure}[t!]
\centering

\begin{tikzpicture}
\begin{axis}[
name = axis1,
xlabel = Elo rating,
x label style = {font=\small},
ylabel = Standard deviation,
y label style = {font=\small},
y tick label style = {/pgf/number format/.cd,fixed,fixed zerofill,precision=2},
width = 0.99\textwidth,
height = 0.6\textwidth,
ymajorgrids = true,
ymin = 0,
legend style = {font=\small,at={(0,-0.15)},anchor=north west,legend columns=1},
legend entries = {{New design, seeding based on UEFA club coefficients ($\sigma_i^n$) $\qquad \qquad \qquad \quad$}, {Old design, seeding based on UEFA club coefficients ($\sigma_i^o$) $\qquad \qquad \qquad \quad \: \:$}, {New design, seeding based on Elo ratings ($\sigma_i^{n, \mathit{Elo}}$) $\qquad \qquad \qquad \qquad \qquad \quad$}, {Old design, seeding based on Elo ratings ($\sigma_i^{o, \mathit{Elo}}$) $\qquad \qquad \qquad \qquad \qquad \quad \:$}, {New design without the play-offs, seeding based on Elo ratings ($\sigma_i^{n, \mathit{Elo}, T16}$)}}
]
\addplot [red, mark=oplus, only marks, mark size=2pt, mark options={solid,thin}] coordinates {
(1898.196899,0.0502465091990149)
(1908.117188,0.0478992598909221)
(1966.385864,0.0303407620346102)
(1918.217041,0.0403000309242126)
(2060.206055,0.00928735371023087)
(1895.175293,0.0540657627959387)
(1861.054688,0.0584048367513852)
(1987.542358,0.024314988333488)
(1950.360352,0.0295049340302962)
(1866.28833,0.0630040857867633)
(1838.463135,0.0618819822786143)
(1917.839355,0.0424219158330095)
(1759.069336,0.0584310907682577)
(1708.752441,0.0475814492245698)
(1833.0625,0.0669668928705875)
(1575.516357,0.0122867275820161)
(1582.821411,0.0148419190659955)
(1748.309204,0.0593388310710017)
(1817.839233,0.0719488231464014)
(1797.253662,0.0634194300124426)
(1674.049072,0.0387321924725897)
(1567.520264,0.0117890609022499)
(1834.71521,0.0605751956606772)
(1553.149902,0.0104810700230858)
(1777.812134,0.0607982612627415)
(1684.736328,0.0432096635722384)
(1653.250366,0.0317833693394292)
(1800.775391,0.0637216642476834)
(1770.44751,0.0632469035353258)
(1727.748535,0.0526139660967355)
(1606.424927,0.020122111304815)
(1791.268799,0.0655045553054577)
};
\addplot [blue, mark=asterisk, only marks, mark size=2.5pt, mark options={solid,semithick}] coordinates {
(1898.196899,0.0856737526661223)
(1908.117188,0.0720836932588351)
(1966.385864,0.0516401981606003)
(1918.217041,0.0673972749573132)
(2060.206055,0.0202774092113446)
(1895.175293,0.0855442873375896)
(1861.054688,0.0955241348788061)
(1987.542358,0.0449403960618751)
(1950.360352,0.0542946058114452)
(1866.28833,0.0935502251179526)
(1838.463135,0.102363197484248)
(1917.839355,0.0724558033800154)
(1759.069336,0.11500793363851)
(1708.752441,0.117003801513321)
(1833.0625,0.106056643713045)
(1575.516357,0.0672746434598702)
(1582.821411,0.0532740996516477)
(1748.309204,0.111212094098951)
(1817.839233,0.126894353128324)
(1797.253662,0.116508411567431)
(1674.049072,0.0850467258722026)
(1567.520264,0.0500089659128408)
(1834.71521,0.115878963980928)
(1553.149902,0.0454536911376627)
(1777.812134,0.129506563250381)
(1684.736328,0.109363571767887)
(1653.250366,0.107745638550368)
(1800.775391,0.139419601578488)
(1770.44751,0.130838045786481)
(1727.748535,0.123062879766306)
(1606.424927,0.084113420257813)
(1791.268799,0.140860195320132)
};
\addplot [orange, mark=otimes, only marks, mark size=2pt, mark options={solid,thin}] coordinates {
(1950.360352,0.016803251866031)
(1898.196899,0.0259508499015083)
(1917.839355,0.021876755322109)
(1908.117188,0.0244802610930354)
(1966.385864,0.0171469525567076)
(1918.217041,0.0214140963508275)
(2060.206055,0.00676614067690436)
(1987.542358,0.0143657218741338)
(1866.28833,0.0307032321338237)
(1838.463135,0.0316093231232228)
(1800.775391,0.0322512167985176)
(1833.0625,0.0314239250104608)
(1817.839233,0.0322578532003153)
(1895.175293,0.0264858069344502)
(1861.054688,0.0299907543310492)
(1834.71521,0.0313879804434168)
(1777.812134,0.0304967631653519)
(1759.069336,0.0303882156048631)
(1708.752441,0.0232643558744757)
(1748.309204,0.0288736973974827)
(1770.44751,0.0305704129698823)
(1797.253662,0.0324065552600677)
(1727.748535,0.0274028460560572)
(1791.268799,0.0316624583737629)
(1684.736328,0.0197062759625535)
(1653.250366,0.016333020497096)
(1582.821411,0.00842453992054598)
(1674.049072,0.019047769647088)
(1567.520264,0.00727562270088861)
(1575.516357,0.00794067694534605)
(1606.424927,0.0110159073259634)
(1553.149902,0.00610496638931081)
}; 
\addplot [ForestGreen, mark=triangle, only marks, mark size=2pt, mark options={solid,thick}] coordinates {
(1950.360352,0.0215044086315014)
(1898.196899,0.0284710512056853)
(1917.839355,0.0254074499608582)
(1908.117188,0.0279177335011968)
(1966.385864,0.0194047804549422)
(1918.217041,0.0257538159120942)
(2060.206055,0.0102105752456386)
(1987.542358,0.0166930451912693)
(1866.28833,0.0429542303493487)
(1838.463135,0.0471571723378629)
(1800.775391,0.0498093422017453)
(1833.0625,0.0492029812548423)
(1817.839233,0.0493892507950272)
(1895.175293,0.0402367905692703)
(1861.054688,0.0412819236618307)
(1834.71521,0.0506334009552222)
(1777.812134,0.0453582280060393)
(1759.069336,0.0497118454553699)
(1708.752441,0.0395013604382998)
(1748.309204,0.046169428110949)
(1770.44751,0.0426191784220364)
(1797.253662,0.050511622424943)
(1727.748535,0.0407389880722282)
(1791.268799,0.0507193830287279)
(1684.736328,0.0259899033294677)
(1653.250366,0.0216139586195405)
(1582.821411,0.0133307263217075)
(1674.049072,0.0243479096578064)
(1567.520264,0.0117713610062703)
(1575.516357,0.0125771564299671)
(1606.424927,0.0161772461900749)
(1553.149902,0.0107398473136409)
};
\addplot [brown, mark=o, only marks, mark size=2pt, mark options={solid,thick}] coordinates {
(1950.360352,0.0225104996300303)
(1898.196899,0.0348882166161966)
(1917.839355,0.0300536040953347)
(1908.117188,0.0330631711165805)
(1966.385864,0.0238438814307084)
(1918.217041,0.0279144871633637)
(2060.206055,0.0100481585123588)
(1987.542358,0.0209888485335133)
(1866.28833,0.039458218345422)
(1838.463135,0.0408471801122565)
(1800.775391,0.0414541834478177)
(1833.0625,0.0407641572035104)
(1817.839233,0.0415807131577455)
(1895.175293,0.0349589182464928)
(1861.054688,0.0388068257620451)
(1834.71521,0.0395762923846956)
(1777.812134,0.0392293897610072)
(1759.069336,0.0394067330491835)
(1708.752441,0.0315155102510873)
(1748.309204,0.0375932208550193)
(1770.44751,0.0389498072843434)
(1797.253662,0.0408646159389249)
(1727.748535,0.0361314283708258)
(1791.268799,0.0410594304981551)
(1684.736328,0.0271277376526343)
(1653.250366,0.0231482369063363)
(1582.821411,0.0132618863148735)
(1674.049072,0.0267083814867476)
(1567.520264,0.0115945974785971)
(1575.516357,0.0125971609286959)
(1606.424927,0.016721090690319)
(1553.149902,0.010224049451523)
};
\end{axis}
\end{tikzpicture}

\captionsetup{justification=centering}
\caption{Standard deviations of qualifying probabilities for the Round of 16 \\ in the old and new UEFA Champions League designs, 2024/25 season}
\label{Fig6}
\end{figure}


Finally, Figure~\ref{Fig6} uncovers standard deviations in five settings as a function of team strengths. The standard deviations are always higher for the middle teams.
The new design substantially decreases the standard deviation if the seeding is inaccurate (i.e., based on UEFA club coefficients), but its advantage is essentially eliminated if the seeding is perfect and the knockout phase play-offs are removed. Furthermore, compared to the old design with inaccurate seeding, the standard deviation is reduced more for weak teams than for strong teams, as the former can occasionally be assigned to a weak group where they would have a reasonable chance to qualify for the Round of 16. Such a favourable schedule is substantially less likely in the new design or if the seeding is accurate. Last but not least, the dots representing the teams lie along a ``smoother'' line in the new design than in the old design if the seeding is inaccurate. Consequently, the new design is fairer in the sense that teams of roughly equal strength are treated more equally.

\subsection{Sensitivity analysis} \label{Sec53}

The pattern seen in Figure~\ref{Fig6} may depend on the distribution of the Elo ratings, the associations of the participating teams, and the difference between the seedings based on UEFA club coefficients and the strength of the teams, which vary across seasons. Therefore, as a robustness check, the Appendix presents analogous figures for the five seasons between 2019/20 and 2023/24, as well as for 2025/26. Figures~\ref{Fig_A1}--\ref{Fig_A6} are essentially similar to Figure~\ref{Fig6}; thus, the results are not driven by the choice of one particular season.

By presenting the 95\% confidence intervals of the standard deviations (see Section~\ref{Sec44} for its calculation), Figure~\ref{Fig_A7} demonstrates that generating 1000 random draws is sufficient for our conclusions above. As expected, the results are less reliable for the middle teams facing the highest draw uncertainty, but the favourable effect of the new design remains clear.

The two less prestigious UEFA club competitions, the UEFA Europa League and the UEFA Conference League, are also organised in an incomplete round-robin format with 36 teams since the 2024/25 season. The format of the UEFA Europa League coincides with that of the UEFA Champions League, discussed in Section~\ref{Sec412}. However, the composition of the teams differs in two important aspects. First, they come from more countries: in the last three seasons, 16 national associations were represented in the Champions League, but their number was 22 or 23 in the Europa League. Therefore, the association-specific restrictions play a smaller role in the Europa League. Second, the variance of Elo ratings is generally higher across all clubs and within the pots in the Champions League than in the Europa League, except for Pot 1 \citep[Table~2]{WinkelmannDeutscherMichels2026}. Consequently, draw uncertainty is expected to be smaller in the Europa League in the incomplete round-robin format, but its advantage compared to the group stage may also be higher as the large variance within Pot 1 implies more imbalanced groups.

The format of the Conference League is different since only six rounds are played instead of eight. In particular, each team is drawn against one opponent from each of the six pots. In addition, Pots 1 and 2, Pots 3 and 4, and Pots 5 and 6 are paired to guarantee that each team plays against one opponent from each pair of pots at home and the other away. The standard deviation of qualifying probabilities is certainly higher in an incomplete round-robin league with fewer rounds, but it should still outperform a comparable group stage with the same number of rounds due to the more representative set of opponents (six instead of only three).

To summarise, the main benefit of the novel tournament design resides in mitigating the influence of inaccurate seeding on draw uncertainty. Hence, although a misaligned seeding could have serious consequences \citep{LaprePalazzolo2022, LaprePalazzolo2023, LapreAmato2025} and UEFA club coefficients have received some criticism from this perspective \citep{Csato2024c}, UEFA should currently be less concerned about using a more accurate seeding (e.g.\ Elo-based), as it matters less in the new design -- just compare the difference between the orange and red, as well as between the green and blue dots in Figure~\ref{Fig6}.

\section{Concluding remarks} \label{Sec6}

For the first time in the tournament design literature, this paper has proposed a new methodology to quantify the uncertainty caused by the draw. The presented approach is based on Monte Carlo simulations and focuses on the variance of qualifying probabilities.
We have used this technique to compare draw uncertainty in the previous group stage and the current incomplete round-robin league phase of the UEFA Champions League, the most prestigious club football tournament in Europe. Since several factors can influence draw uncertainty, we have distinguished three channels (first stage, play-off, seeding) in order to decompose the effects of the 2024/25 reform for each team.

Our results show that the main advantage of an incomplete round-robin tournament is the insensitivity of its draw uncertainty to the seeding of the teams. This can be especially relevant for tournament organisers if they are unsure about the strength of the participants. On the other hand, if the strength of the teams can be reliably estimated, draw uncertainty does not differ much in the two formats. Hence, the benefits of the innovative incomplete round-robin format depend on outcome uncertainty, which is quite different across sports and leagues. For example, higher scoring rates tend to decrease outcome uncertainty \citep{ScarfParmaMcHale2019}, potentially allowing for a more accurate seeding of the teams.


\section*{Data availability}
\addcontentsline{toc}{section}{Data availability}

Our data, code and results are publicly available on GitHub: \url{https://github.com/Gyimesiandras/uncertainty_of_draws}.

\section*{Acknowledgements}
\addcontentsline{toc}{section}{Acknowledgements}

We thank the Editor-in-Chief \emph{Mike Yearworth} and four anonymous reviewers for their insightful comments that helped shape our paper.

The research was supported by the National Research, Development and Innovation Office under Grants Advanced 152220 and FK 145838, and by the J\'anos Bolyai Research Scholarship of the Hungarian Academy of Sciences. \\
This research was also supported by the NWO Gravitation project NETWORKS under grant no.~024.002.003, and the Special Research Fund of Ghent University under grant no.~BOF/24J/2021/188.

\bibliographystyle{apalike} 
\bibliography{All_references}

\clearpage
\section*{Appendix}
\addcontentsline{toc}{section}{Appendix}


Here, we discuss an integer program proposed in \citet{DevriesereGoossensSpieksma2026}, which is able to check whether the UEFA Champions League league phase draw can be completed when a particular match is drawn. Let $T$ denote the set of teams, let $C$ denote the set of participating associations, and let $T_c$ denote the set of teams of association $c \in C$. We use $a(i)$ to denote the association of team $i \in T$.
Next, $P$ denotes the set of pots, and $P_h$ denotes the set of teams in pot $h \in P$. For all team pairs $i, j \in T$, the binary variable $m_{ij}$ equals 1 if team $i$ plays at home against team $j$, and 0 otherwise. Parameter $\omega_{ij}$ is 1 if teams $i$ and $j$ were matched in a previous iteration, and 0 otherwise.


Suppose that, at a given point of the draw procedure, the match between teams $u \in T$ (home) and $v \in T$ (away) is proposed. This match is added to the set of previously drawn matches only if a feasible solution exists to the following model:
\begin{align}
    & \sum_{j \in T_{a(i)}}(m_{ij}+m_{ji}) = 0 & \forall i \in T \label{draw:c8} \\
    & \sum_{j \in T_c}(m_{ij} + m_{ji}) \leq 2 & \forall \ i \in T, \forall \ c \in C: a(i) \neq c \label{draw:c1} \\
    & \sum_{j \in P_h}m_{ij} = 1 & \forall \ i \in T, \forall \ h = 1,\dots,|P| \label{draw:c3} \\
    & \sum_{j \in P_h}m_{ji} = 1 & \forall \ i \in T, \forall \ h = 1,\dots,|P| \label{draw:c4} \\
    & m_{ij} + m_{ji} \leq 1 & \forall \ i,j \in T \label{draw:c2} \\
    & m_{ij} \geq \omega_{ij} & \forall \ i, j \in T \label{draw:c5}\\
    & m_{ii} = 0 & \forall \ i \in T \label{draw:c6} \\
    & m_{uv} = 1 & \label{draw:c9}\\
    & m_{ij} \in \{0, 1\} & \forall \ i, j \in T. \label{draw:c7}
\end{align}

Constraints~\eqref{draw:c8} ensure that no team plays against a team from the same association. Constraints~\eqref{draw:c1} ensure that each team is matched with at most two teams from the same (but different from its own) association. Constraints~\eqref{draw:c3} and \eqref{draw:c4} guarantee that a team faces exactly one team from each pot at home and one team from each pot away. According to constraints~\eqref{draw:c2}, no team is matched to the same opponent twice. The previously drawn matches are fixed by constraints~\eqref{draw:c5}. Naturally, a team cannot play against itself, which is stated by constraints~\eqref{draw:c6}.
Finally, constraint~\eqref{draw:c9} implies that team $u$ plays at home against team $v$, and constraints~\eqref{draw:c7} specify that the variables are binary.

If the integer program given by constraints~\eqref{draw:c8}--\eqref{draw:c7} has a feasible solution, we accept the drawn match between teams $u$ and $v$, update the value for the corresponding parameter $\omega_{uv}$ to 1, and continue the draw with the next pair of teams.

However, if formulation~\eqref{draw:c8}--\eqref{draw:c7} has no feasible solution, then $m_{uv} = 0$ still ensures the existence of a feasible solution (otherwise, the previously drawn matches have already lead to infeasibility).
Hence, we remove team $v$ from the set of opponents available for team $u$, and draw a team from this reduced set of possible opponents. The set of opponents available for team $u$ never becomes empty.

The presented approach mimics the official iterative procedure of UEFA, which proceeds from Pot 1 to Pot 4 such that all the opponents of a team are drawn before the next team is drawn. Naturally, all draws from the pots and the sets of possible opponents are random. This procedure guarantees that a feasible draw is found if one exists.

\clearpage
\setcounter{figure}{0}
\renewcommand{\thefigure}{A.\arabic{figure}}

\begin{figure}[ht!]
\centering

\begin{tikzpicture}
\begin{axis}[
name = axis1,
xlabel = Elo rating,
x label style = {font=\small},
ylabel = Standard deviation,
y label style = {font=\small},
y tick label style = {/pgf/number format/.cd,fixed,fixed zerofill,precision=2},
width = 0.99\textwidth,
height = 0.6\textwidth,
ymajorgrids = true,
ymin = 0,
legend style = {font=\small,at={(0,-0.15)},anchor=north west,legend columns=1},
legend entries = {{New design, seeding based on UEFA club coefficients ($\sigma_i^n$) $\qquad \qquad \qquad \quad$}, {Old design, seeding based on UEFA club coefficients ($\sigma_i^o$) $\qquad \qquad \qquad \quad \: \:$}, {New design, seeding based on Elo ratings ($\sigma_i^{n, \mathit{Elo}}$) $\qquad \qquad \qquad \qquad \qquad \quad$}, {Old design, seeding based on Elo ratings ($\sigma_i^{o, \mathit{Elo}}$) $\qquad \qquad \qquad \qquad \qquad \quad \:$}, {New design without the play-offs, seeding based on Elo ratings ($\sigma_i^{n, \mathit{Elo}, T16}$)}}
]
\addplot [red, mark=oplus, only marks, mark size=2pt, mark options={solid,thin}] coordinates {
(1830.73461914,0.0496267308253567)
(2065.00097656,0.00596590244573208)
(1817.74121094,0.047407965429727)
(2050.77294922,0.00710511919102831)
(1892.83532715,0.030409484832331)
(1734.38977051,0.0433298280402542)
(1950.14770508,0.0244251010348076)
(1868.35717773,0.0432247499704742)
(1985.17431641,0.0170716171205633)
(1888.0703125,0.0385515735583364)
(1875.74890137,0.0444055332969596)
(1779.5255127,0.0482217213028051)
(1682.77893066,0.0325881689076381)
(1680.7355957,0.0294278716632826)
(1655.33117676,0.0243699013352761)
(1809.71508789,0.0467503290819355)
(1688.03491211,0.0329627979098566)
(1844.27880859,0.0456342362982282)
(1752.73132324,0.0477678427165996)
(1845.15441895,0.0389673982853504)
(1760.89599609,0.0475741429147607)
(1773.42370605,0.0484518034055829)
(1806.3470459,0.0529636337735856)
(1635.61804199,0.019886706919752)
(1799.64257812,0.0517554661100861)
(1683.83520508,0.0311066746729574)
(1590.10253906,0.0112824673307335)
(1731.33740234,0.0431640160699081)
(1800.66516113,0.0484466210492651)
(1717.47033691,0.0399218162840246)
(1672.00622559,0.0255754781007136)
(1621.72827148,0.0153727322521696)
};
\addplot [blue, mark=asterisk, only marks, mark size=2.5pt, mark options={solid,semithick}] coordinates {
(1830.73461914,0.077366713303121)
(2065.00097656,0.0124699745093671)
(1817.74121094,0.0720612787764615)
(2050.77294922,0.013752253650965)
(1892.83532715,0.0527054697877321)
(1734.38977051,0.091257480256156)
(1950.14770508,0.0326225100837469)
(1868.35717773,0.0737083952581491)
(1985.17431641,0.0255825700798784)
(1888.0703125,0.052616255913799)
(1875.74890137,0.061751598229389)
(1779.5255127,0.0909951322274389)
(1682.77893066,0.0679195093014058)
(1680.7355957,0.0602936348332177)
(1655.33117676,0.0517165083240664)
(1809.71508789,0.082786633810886)
(1688.03491211,0.0701101994796126)
(1844.27880859,0.0648228758713792)
(1752.73132324,0.0845969675929695)
(1845.15441895,0.0660331346436838)
(1760.89599609,0.0803824084460499)
(1773.42370605,0.0890007253909771)
(1806.3470459,0.0848513852661321)
(1635.61804199,0.0430681813319676)
(1799.64257812,0.0827353583089752)
(1683.83520508,0.0569686670813793)
(1590.10253906,0.031477454251037)
(1731.33740234,0.0869356085619468)
(1800.66516113,0.0925718749028182)
(1717.47033691,0.0795964830322993)
(1672.00622559,0.0641583913309723)
(1621.72827148,0.0271866925581663)
};
\addplot [orange, mark=otimes, only marks, mark size=2pt, mark options={solid,thin}] coordinates {
(1845.15441895,0.0252765635255994)
(1985.17431641,0.0145534362439427)
(2065.00097656,0.00518424656169172)
(2050.77294922,0.00614036677276726)
(1950.14770508,0.0191413202559578)
(1888.0703125,0.0276635144917278)
(1892.83532715,0.0206847571457385)
(1817.74121094,0.0340427441230496)
(1799.64257812,0.0334816579251598)
(1868.35717773,0.0301571703899077)
(1717.47033691,0.0247414433293213)
(1688.03491211,0.020803624246313)
(1731.33740234,0.0269263740711708)
(1844.27880859,0.0321471846431454)
(1779.5255127,0.0314709261731236)
(1752.73132324,0.0300669043091359)
(1773.42370605,0.0330836044572517)
(1672.00622559,0.0184671269768413)
(1875.74890137,0.0291203639330898)
(1809.71508789,0.0342186842873065)
(1682.77893066,0.0197973965942977)
(1800.66516113,0.0318903511190218)
(1806.3470459,0.0336355209900527)
(1655.33117676,0.015950370350381)
(1760.89599609,0.0315097156632868)
(1734.38977051,0.0280238893513084)
(1830.73461914,0.0340718689133903)
(1635.61804199,0.0136452244815229)
(1683.83520508,0.0193350650565428)
(1590.10253906,0.00853650372786751)
(1621.72827148,0.0116812416894527)
(1680.7355957,0.0202859584715909)
}; 
\addplot [ForestGreen, mark=triangle, only marks, mark size=2pt, mark options={solid,thick}] coordinates {
(1845.15441895,0.0403191390902258)
(1985.17431641,0.0127399806163944)
(2065.00097656,0.00763506733783359)
(2050.77294922,0.00844226840937292)
(1950.14770508,0.0169516310954668)
(1888.0703125,0.0250493847364696)
(1892.83532715,0.0241985427137559)
(1817.74121094,0.0476782744744067)
(1799.64257812,0.0484981020443264)
(1868.35717773,0.0292124328023229)
(1717.47033691,0.0344983203809854)
(1688.03491211,0.0317033723729798)
(1731.33740234,0.0376948308959767)
(1844.27880859,0.0453033285264513)
(1779.5255127,0.0428788809839347)
(1752.73132324,0.0405975191182732)
(1773.42370605,0.0444919033697758)
(1672.00622559,0.0269196832884109)
(1875.74890137,0.0305863930275938)
(1809.71508789,0.0485709494258464)
(1682.77893066,0.0278414902542874)
(1800.66516113,0.0493710835904469)
(1806.3470459,0.0525415164127088)
(1655.33117676,0.0232848763033963)
(1760.89599609,0.0402719414514407)
(1734.38977051,0.0378913487528527)
(1830.73461914,0.0478596788524522)
(1635.61804199,0.0184250287838496)
(1683.83520508,0.0285001159227047)
(1590.10253906,0.0150003141775573)
(1621.72827148,0.0185620931551994)
(1680.7355957,0.0289773842815358)
};
\addplot [brown, mark=o, only marks, mark size=2pt, mark options={solid,thick}] coordinates {
(1845.15441895,0.0313533635900266)
(1985.17431641,0.0220449735054696)
(2065.00097656,0.0076889712757481)
(2050.77294922,0.00910876202046938)
(1950.14770508,0.0267986714743995)
(1888.0703125,0.0377456031155227)
(1892.83532715,0.0263894992311273)
(1817.74121094,0.0438265311864381)
(1799.64257812,0.0420940332069954)
(1868.35717773,0.0392512238991038)
(1717.47033691,0.0322261644735559)
(1688.03491211,0.0274215659341015)
(1731.33740234,0.0348646804939604)
(1844.27880859,0.0402847224001713)
(1779.5255127,0.039808927009776)
(1752.73132324,0.0379320170425581)
(1773.42370605,0.0418893840548712)
(1672.00622559,0.0235864642601704)
(1875.74890137,0.0388744113635031)
(1809.71508789,0.0422997198844425)
(1682.77893066,0.026359761778913)
(1800.66516113,0.0402508211046632)
(1806.3470459,0.0428340817288939)
(1655.33117676,0.021993860072322)
(1760.89599609,0.0386844429988352)
(1734.38977051,0.0366678170581207)
(1830.73461914,0.0436295962124985)
(1635.61804199,0.018260784628686)
(1683.83520508,0.0245839592020905)
(1590.10253906,0.012150733283194)
(1621.72827148,0.0172613694089975)
(1680.7355957,0.0266732723800335)
};
\end{axis}
\end{tikzpicture}

\captionsetup{justification=centering}
\caption{Standard deviations of qualifying probabilities for the Round of 16 \\ in the old and new UEFA Champions League designs, 2019/20 season}
\label{Fig_A1}
\end{figure}

\begin{figure}[ht!]
\centering

\begin{tikzpicture}
\begin{axis}[
name = axis1,
xlabel = Elo rating,
x label style = {font=\small},
ylabel = Standard deviation,
y label style = {font=\small},
y tick label style = {/pgf/number format/.cd,fixed,fixed zerofill,precision=2},
width = 0.99\textwidth,
height = 0.6\textwidth,
ymajorgrids = true,
ymin = 0,
legend style = {font=\small,at={(0,-0.15)},anchor=north west,legend columns=1},
legend entries = {{New design, seeding based on UEFA club coefficients ($\sigma_i^n$) $\qquad \qquad \qquad \quad$}, {Old design, seeding based on UEFA club coefficients ($\sigma_i^o$) $\qquad \qquad \qquad \quad \: \:$}, {New design, seeding based on Elo ratings ($\sigma_i^{n, \mathit{Elo}}$) $\qquad \qquad \qquad \qquad \qquad \quad$}, {Old design, seeding based on Elo ratings ($\sigma_i^{o, \mathit{Elo}}$) $\qquad \qquad \qquad \qquad \qquad \quad \:$}, {New design without the play-offs, seeding based on Elo ratings ($\sigma_i^{n, \mathit{Elo}, T16}$)}}
]
\addplot [red, mark=oplus, only marks, mark size=2pt, mark options={solid,thin}] coordinates {
(1919.92993164,0.0354934329670121)
(1890.39416504,0.0422543259212481)
(1991.63098145,0.0141638776975597)
(2044.85327148,0.00932751856618912)
(1849.34191895,0.0447379462518658)
(1825.86865234,0.0507921426773646)
(1841.54016113,0.0474282882521236)
(1765.77270508,0.0501318255795252)
(1862.48913574,0.0491558440880432)
(1954.91345215,0.0269872736106658)
(1926.95141602,0.0332586116855658)
(1837.94238281,0.0508915122296146)
(1738.37231445,0.0500766733041364)
(1967.77905273,0.0167792445314754)
(1767.00500488,0.0530582015974605)
(1721.25195312,0.0392607203597077)
(1745.82849121,0.051221073684465)
(1671.48840332,0.033906584205025)
(1601.2109375,0.0168299306464871)
(1860.72937012,0.0418083524949063)
(1765.83337402,0.0543895081205336)
(1818.56286621,0.0466781071776662)
(1771.40185547,0.0551483568228489)
(1533.8861084,0.00790280699235141)
(1840.02722168,0.0486407766812595)
(1701.66992188,0.0403915849668732)
(1729.00915527,0.046519039080757)
(1669.45349121,0.0327839934453725)
(1586.30383301,0.0122905332759896)
(1590.16479492,0.012912488968877)
(1581.24560547,0.013653787200074)
(1580.03295898,0.0137364872541011)
};
\addplot [blue, mark=asterisk, only marks, mark size=2.5pt, mark options={solid,semithick}] coordinates {
(1919.92993164,0.0622240592097988)
(1890.39416504,0.0758089851748012)
(1991.63098145,0.0443233681139076)
(2044.85327148,0.0250676665029656)
(1849.34191895,0.0929311309039715)
(1825.86865234,0.105123381515918)
(1841.54016113,0.103082650571514)
(1765.77270508,0.10339273108319)
(1862.48913574,0.0840385377956735)
(1954.91345215,0.0510311059038482)
(1926.95141602,0.0641969778335629)
(1837.94238281,0.0917102150174044)
(1738.37231445,0.0899989998331702)
(1967.77905273,0.0285852604871307)
(1767.00500488,0.0972774825929394)
(1721.25195312,0.107061308991077)
(1745.82849121,0.123962771704931)
(1671.48840332,0.0698137589371675)
(1601.2109375,0.048930027371529)
(1860.72937012,0.0956756158457706)
(1765.83337402,0.111428494472376)
(1818.56286621,0.110003213912463)
(1771.40185547,0.120282271681578)
(1533.8861084,0.0271630917340525)
(1840.02722168,0.0981551420757767)
(1701.66992188,0.0785873137655492)
(1729.00915527,0.0844996057231474)
(1669.45349121,0.0661483857167575)
(1586.30383301,0.0393697347592667)
(1590.16479492,0.0307436935108762)
(1581.24560547,0.0389565142948648)
(1580.03295898,0.0380279436664606)
};
\addplot [orange, mark=otimes, only marks, mark size=2pt, mark options={solid,thin}] coordinates {
(1818.56286621,0.029513103179084)
(1954.91345215,0.0167152900040628)
(1991.63098145,0.010126072107996)
(1926.95141602,0.0171729936444031)
(1860.72937012,0.0240247995828268)
(2044.85327148,0.0061002724795147)
(1919.92993164,0.0204126135953678)
(1840.02722168,0.0277717959204775)
(1825.86865234,0.0297752023280057)
(1771.40185547,0.029263216755081)
(1967.77905273,0.0122621737430565)
(1841.54016113,0.0277390297734577)
(1765.77270508,0.0282009739891007)
(1729.00915527,0.0268456857212075)
(1701.66992188,0.0220831025768441)
(1837.94238281,0.0278557685918706)
(1767.00500488,0.0290649013845307)
(1721.25195312,0.0271777999258367)
(1586.30383301,0.00920764659159743)
(1849.34191895,0.0263622807405931)
(1862.48913574,0.0272410806774708)
(1601.2109375,0.0102417229453295)
(1765.83337402,0.0303567312326464)
(1580.03295898,0.00883439854192676)
(1738.37231445,0.028290436189281)
(1890.39416504,0.0239442988715153)
(1671.48840332,0.0175617821188851)
(1581.24560547,0.0087176600987649)
(1669.45349121,0.0178622433848254)
(1590.16479492,0.00993640191835471)
(1533.8861084,0.0057920835829805)
(1745.82849121,0.0280643438975223)
}; 
\addplot [ForestGreen, mark=triangle, only marks, mark size=2pt, mark options={solid,thick}] coordinates {
(1818.56286621,0.0424676615470793)
(1954.91345215,0.0172521190179818)
(1991.63098145,0.0126240584144264)
(1926.95141602,0.0187194054430193)
(1860.72937012,0.0377813100141642)
(2044.85327148,0.0087653352532634)
(1919.92993164,0.0210825309702633)
(1840.02722168,0.0371605885359791)
(1825.86865234,0.0428196522945087)
(1771.40185547,0.0455316839711123)
(1967.77905273,0.014471418242382)
(1841.54016113,0.0407665231459516)
(1765.77270508,0.0468863038048905)
(1729.00915527,0.0499469355954434)
(1701.66992188,0.0390142180905015)
(1837.94238281,0.0414036791680238)
(1767.00500488,0.0467510252546654)
(1721.25195312,0.043650635929125)
(1586.30383301,0.0145335084156189)
(1849.34191895,0.035531848196178)
(1862.48913574,0.0305657220324232)
(1601.2109375,0.0160195574914609)
(1765.83337402,0.0474383947588585)
(1580.03295898,0.0137870961416833)
(1738.37231445,0.0427129262644234)
(1890.39416504,0.0268668863209077)
(1671.48840332,0.025492187089432)
(1581.24560547,0.0135307898244438)
(1669.45349121,0.0244442422050056)
(1590.16479492,0.0146837475953329)
(1533.8861084,0.00985603878343123)
(1745.82849121,0.0450042237479457)
};
\addplot [brown, mark=o, only marks, mark size=2pt, mark options={solid,thick}] coordinates {
(1818.56286621,0.037510615100865)
(1954.91345215,0.0235566639492186)
(1991.63098145,0.0145713553039995)
(1926.95141602,0.0221485298241976)
(1860.72937012,0.0312929027858517)
(2044.85327148,0.00890607445298781)
(1919.92993164,0.0272499288553681)
(1840.02722168,0.0346570746576197)
(1825.86865234,0.0378638091459779)
(1771.40185547,0.037523059309976)
(1967.77905273,0.0165528628420666)
(1841.54016113,0.0354237184783138)
(1765.77270508,0.0362045880545021)
(1729.00915527,0.0341260497460704)
(1701.66992188,0.0286756761978755)
(1837.94238281,0.0354840204416961)
(1767.00500488,0.0372461863067133)
(1721.25195312,0.0351918283665936)
(1586.30383301,0.0140555497038528)
(1849.34191895,0.0329378142961947)
(1862.48913574,0.0352050344475112)
(1601.2109375,0.0152396893712481)
(1765.83337402,0.0384882940616985)
(1580.03295898,0.0124209255045826)
(1738.37231445,0.0361829240802204)
(1890.39416504,0.0317737640685375)
(1671.48840332,0.0232112098836804)
(1581.24560547,0.0128101805110125)
(1669.45349121,0.0231698621815822)
(1590.16479492,0.0147872816873048)
(1533.8861084,0.00814676382134408)
(1745.82849121,0.0353418692023882)
};
\end{axis}
\end{tikzpicture}

\captionsetup{justification=centering}
\caption{Standard deviations of qualifying probabilities for the Round of 16 \\ in the old and new UEFA Champions League designs, 2020/21 season}
\label{Fig_A2}
\end{figure}

\begin{figure}[ht!]
\centering

\begin{tikzpicture}
\begin{axis}[
name = axis1,
xlabel = Elo rating,
x label style = {font=\small},
ylabel = Standard deviation,
y label style = {font=\small},
y tick label style = {/pgf/number format/.cd,fixed,fixed zerofill,precision=2},
width = 0.99\textwidth,
height = 0.6\textwidth,
ymajorgrids = true,
ymin = 0,
legend style = {font=\small,at={(0,-0.15)},anchor=north west,legend columns=1},
legend entries = {{New design, seeding based on UEFA club coefficients ($\sigma_i^n$) $\qquad \qquad \qquad \quad$}, {Old design, seeding based on UEFA club coefficients ($\sigma_i^o$) $\qquad \qquad \qquad \quad \: \:$}, {New design, seeding based on Elo ratings ($\sigma_i^{n, \mathit{Elo}}$) $\qquad \qquad \qquad \qquad \qquad \quad$}, {Old design, seeding based on Elo ratings ($\sigma_i^{o, \mathit{Elo}}$) $\qquad \qquad \qquad \qquad \qquad \quad \:$}, {New design without the play-offs, seeding based on Elo ratings ($\sigma_i^{n, \mathit{Elo}, T16}$)}}
]
\addplot [red, mark=oplus, only marks, mark size=2pt, mark options={solid,thin}] coordinates {
(1898.23266602,0.049112280084059)
(1767.49645996,0.0611763573425036)
(1932.66455078,0.0335484580127766)
(1938.4901123,0.0324493433505808)
(1814.10266113,0.051867301191916)
(1820.1932373,0.0605051423289645)
(1992.09020996,0.0212505129420516)
(1850.68981934,0.0568863817132398)
(1834.9576416,0.0496377404832279)
(1889.40881348,0.0500950639727552)
(1819.03417969,0.0612431317388749)
(1805.97973633,0.0630653351731401)
(1916.14477539,0.0428786947850149)
(2012.69165039,0.0148492171595088)
(1785.54309082,0.059639040897714)
(1829.92297363,0.0491267768045004)
(1947.64782715,0.0280128916232705)
(1609.77856445,0.0173631295931715)
(1769.11657715,0.0594209084696907)
(1906.73071289,0.0354292612334995)
(1727.00830078,0.0497089605177553)
(1804.71252441,0.0621292134036651)
(1941.68847656,0.0346399848809516)
(1844.78210449,0.0574759983504975)
(1689.9342041,0.0382012121353316)
(1673.37390137,0.0342999800675253)
(1672.56018066,0.0331823823964088)
(1585.4296875,0.013882580813741)
(1562.56469727,0.00981320615369997)
(1727.37231445,0.0520496292004074)
(1688.97680664,0.0384530174830735)
(1429.40612793,0.00184222088191883)
};
\addplot [blue, mark=asterisk, only marks, mark size=2.5pt, mark options={solid,semithick}] coordinates {
(1898.23266602,0.0900595281945149)
(1767.49645996,0.104513074333821)
(1932.66455078,0.0759733892548889)
(1938.4901123,0.0535620936192087)
(1814.10266113,0.0918499951693663)
(1820.1932373,0.127547806888406)
(1992.09020996,0.0478862643177593)
(1850.68981934,0.122308459622772)
(1834.9576416,0.0978887883973129)
(1889.40881348,0.088870278081743)
(1819.03417969,0.1082946050248)
(1805.97973633,0.120521895867031)
(1916.14477539,0.0843780882740346)
(2012.69165039,0.0288899853799765)
(1785.54309082,0.102450060472851)
(1829.92297363,0.0981867731286522)
(1947.64782715,0.0685969405385638)
(1609.77856445,0.044679209883298)
(1769.11657715,0.104731979349664)
(1906.73071289,0.0841060381700188)
(1727.00830078,0.084326455487558)
(1804.71252441,0.132010692377178)
(1941.68847656,0.068554413012902)
(1844.78210449,0.112353648994781)
(1689.9342041,0.122320655417398)
(1673.37390137,0.0772015242034724)
(1672.56018066,0.0591002499448483)
(1585.4296875,0.0411305222901906)
(1562.56469727,0.0468411879961738)
(1727.37231445,0.0945969056024009)
(1688.97680664,0.0764592614250076)
(1429.40612793,0.0113886160957588)
};
\addplot [orange, mark=otimes, only marks, mark size=2pt, mark options={solid,thin}] coordinates {
(1992.09020996,0.0123612892722593)
(1906.73071289,0.0243588625593476)
(1805.97973633,0.0328416773592911)
(2012.69165039,0.0104315932214181)
(1938.4901123,0.0204349501390923)
(1785.54309082,0.0331736773341134)
(1898.23266602,0.026613430043094)
(1947.64782715,0.0178693306247598)
(1844.78210449,0.0325088059834019)
(1850.68981934,0.0326871375308386)
(1889.40881348,0.0261279437469626)
(1769.11657715,0.0332071629572849)
(1932.66455078,0.019708900521332)
(1820.1932373,0.0330501179164961)
(1941.68847656,0.0201285562019429)
(1672.56018066,0.0198609369319732)
(1814.10266113,0.0336395412384702)
(1689.9342041,0.0234854100131014)
(1673.37390137,0.020160739356765)
(1829.92297363,0.0335091755919009)
(1834.9576416,0.0334442948327044)
(1916.14477539,0.023528237012937)
(1804.71252441,0.0354194253862862)
(1609.77856445,0.0123508172477244)
(1688.97680664,0.022489610191521)
(1585.4296875,0.00937766386578126)
(1562.56469727,0.00732105004584546)
(1819.03417969,0.0343336319413237)
(1727.37231445,0.0286959951759673)
(1767.49645996,0.0323429844733699)
(1727.00830078,0.0278271585101812)
(1429.40612793,0.00177394218363251)
}; 
\addplot [ForestGreen, mark=triangle, only marks, mark size=2pt, mark options={solid,thick}] coordinates {
(1992.09020996,0.0200714838630447)
(1906.73071289,0.0332443773836663)
(1805.97973633,0.0513393623853095)
(2012.69165039,0.0158115141267046)
(1938.4901123,0.0282424930452928)
(1785.54309082,0.0500753734087128)
(1898.23266602,0.0425334493112222)
(1947.64782715,0.0265013970017251)
(1844.78210449,0.0525661492221377)
(1850.68981934,0.0494468491907714)
(1889.40881348,0.0387731186613965)
(1769.11657715,0.04636505880162)
(1932.66455078,0.0285784255235555)
(1820.1932373,0.0542748053012675)
(1941.68847656,0.0272392894527352)
(1672.56018066,0.024997195418262)
(1814.10266113,0.0507603377736091)
(1689.9342041,0.0273275095238622)
(1673.37390137,0.0252362282882807)
(1829.92297363,0.0502714371526555)
(1834.9576416,0.0480665536270032)
(1916.14477539,0.0307930673528474)
(1804.71252441,0.0470838209706604)
(1609.77856445,0.0149235950323201)
(1688.97680664,0.0270176760440053)
(1585.4296875,0.0138487964879322)
(1562.56469727,0.0110224845731879)
(1819.03417969,0.0538455586207737)
(1727.37231445,0.0440812380900486)
(1767.49645996,0.0456944290351115)
(1727.00830078,0.0428101060636526)
(1429.40612793,0.0049337407509716)
};
\addplot [brown, mark=o, only marks, mark size=2pt, mark options={solid,thick}] coordinates {
(1992.09020996,0.0169889846842182)
(1906.73071289,0.0326261482487521)
(1805.97973633,0.0413457425792479)
(2012.69165039,0.0141617748487563)
(1938.4901123,0.0261920862357579)
(1785.54309082,0.0426148351536783)
(1898.23266602,0.0343912711968391)
(1947.64782715,0.023744876819541)
(1844.78210449,0.0402584626484273)
(1850.68981934,0.0416348846836715)
(1889.40881348,0.0327532525186515)
(1769.11657715,0.0419703104940023)
(1932.66455078,0.0267282902839202)
(1820.1932373,0.0420401135363137)
(1941.68847656,0.026707811182839)
(1672.56018066,0.027121174423356)
(1814.10266113,0.0430537327406452)
(1689.9342041,0.0321701113934513)
(1673.37390137,0.0286117357464105)
(1829.92297363,0.0423706462974978)
(1834.9576416,0.0428375049089793)
(1916.14477539,0.0306413215541789)
(1804.71252441,0.044091757439164)
(1609.77856445,0.0184600870478396)
(1688.97680664,0.0306460716194704)
(1585.4296875,0.0152844361685012)
(1562.56469727,0.0122220568921387)
(1819.03417969,0.0432035357523143)
(1727.37231445,0.0373539676996019)
(1767.49645996,0.0401775335006792)
(1727.00830078,0.0364056599456194)
(1429.40612793,0.00308566773087209)
};
\end{axis}
\end{tikzpicture}

\captionsetup{justification=centering}
\caption{Standard deviations of qualifying probabilities for the Round of 16 \\ in the old and new UEFA Champions League designs, 2021/22 season}
\label{Fig_A3}
\end{figure}

\begin{figure}[ht!]
\centering

\begin{tikzpicture}
\begin{axis}[
name = axis1,
xlabel = Elo rating,
x label style = {font=\small},
ylabel = Standard deviation,
y label style = {font=\small},
y tick label style = {/pgf/number format/.cd,fixed,fixed zerofill,precision=2},
width = 0.99\textwidth,
height = 0.6\textwidth,
ymajorgrids = true,
ymin = 0,
legend style = {font=\small,at={(0,-0.15)},anchor=north west,legend columns=1},
legend entries = {{New design, seeding based on UEFA club coefficients ($\sigma_i^n$) $\qquad \qquad \qquad \quad$}, {Old design, seeding based on UEFA club coefficients ($\sigma_i^o$) $\qquad \qquad \qquad \quad \: \:$}, {New design, seeding based on Elo ratings ($\sigma_i^{n, \mathit{Elo}}$) $\qquad \qquad \qquad \qquad \qquad \quad$}, {Old design, seeding based on Elo ratings ($\sigma_i^{o, \mathit{Elo}}$) $\qquad \qquad \qquad \qquad \qquad \quad \:$}, {New design without the play-offs, seeding based on Elo ratings ($\sigma_i^{n, \mathit{Elo}, T16}$)}}
]
\addplot [red, mark=oplus, only marks, mark size=2pt, mark options={solid,thin}] coordinates {
(1857.6895752,0.0420438732549497)
(2031.7364502,0.0112365847496842)
(1799.16796875,0.0577600115818716)
(1800.84069824,0.0578699969980348)
(2013.15661621,0.0139194657773755)
(1723.5,0.0458451870120755)
(1886.71411133,0.0416498392326802)
(1621.82702637,0.01693012995521)
(1880.93579102,0.0435551200719787)
(1993.74035645,0.0192343348113636)
(1741.96362305,0.0490860198156394)
(1804.87023926,0.0471652874510518)
(1966.92700195,0.0252862090351118)
(1873.86962891,0.0468280962093762)
(1793.28295898,0.0578076451937185)
(1885.56103516,0.0476278743323005)
(1876.80871582,0.047278694503486)
(1603.12768555,0.0135717229069364)
(1868.44873047,0.0527602206200835)
(1806.69091797,0.0561693602244856)
(1837.83227539,0.0460636479087215)
(1747.40930176,0.0484225416675079)
(1820.65563965,0.0543645250193596)
(1768.06152344,0.0534079259806778)
(1789.2902832,0.0467960475795667)
(1568.34191895,0.00864241646545469)
(1647.30444336,0.0217994747895701)
(1739.10449219,0.0477225630810734)
(1730.24450684,0.0477769480059239)
(1628.15588379,0.0190654903929567)
(1637.4041748,0.0185381509066854)
(1672.93603516,0.0302696012487737)
};
\addplot [blue, mark=asterisk, only marks, mark size=2.5pt, mark options={solid,semithick}] coordinates {
(1857.6895752,0.0890058153783851)
(2031.7364502,0.0199268568425535)
(1799.16796875,0.0967773260730107)
(1800.84069824,0.0948288314927601)
(2013.15661621,0.0249265912421043)
(1723.5,0.10996568979428)
(1886.71411133,0.0776167967784174)
(1621.82702637,0.0432761707057239)
(1880.93579102,0.0818585651944602)
(1993.74035645,0.032277207160814)
(1741.96362305,0.0814148835426439)
(1804.87023926,0.0958118896442252)
(1966.92700195,0.0393918216844006)
(1873.86962891,0.0683171283136303)
(1793.28295898,0.110958847510421)
(1885.56103516,0.0679317080306094)
(1876.80871582,0.0685566623454785)
(1603.12768555,0.0386514036226629)
(1868.44873047,0.0857408105081039)
(1806.69091797,0.0967376489466419)
(1837.83227539,0.0841042295097067)
(1747.40930176,0.105726429211592)
(1820.65563965,0.0937594634582984)
(1768.06152344,0.127006193093424)
(1789.2902832,0.0713206450434193)
(1568.34191895,0.0291558380756564)
(1647.30444336,0.0493367042006393)
(1739.10449219,0.114216817151943)
(1730.24450684,0.114087098613603)
(1628.15588379,0.0464073040465493)
(1637.4041748,0.0614486153084292)
(1672.93603516,0.0629417951251279)
};
\addplot [orange, mark=otimes, only marks, mark size=2pt, mark options={solid,thin}] coordinates {
(2013.15661621,0.0100769407925092)
(1886.71411133,0.0279185860574537)
(1868.44873047,0.0301559495339194)
(1820.65563965,0.0325551334310384)
(1747.40930176,0.028279320075701)
(1876.80871582,0.0274628455168776)
(1966.92700195,0.0175441670591054)
(1993.74035645,0.0134221999482742)
(1880.93579102,0.0251262430954588)
(1885.56103516,0.0271219541404449)
(1730.24450684,0.0252256446726174)
(2031.7364502,0.00804545619373851)
(1800.84069824,0.032658982732981)
(1768.06152344,0.0312645657745805)
(1873.86962891,0.0266454455915222)
(1857.6895752,0.0305780610375612)
(1793.28295898,0.0307793236729751)
(1723.5,0.0227401476097521)
(1806.69091797,0.0326993018938378)
(1672.93603516,0.0172721168946375)
(1799.16796875,0.0315283855453319)
(1621.82702637,0.0107049070072577)
(1804.87023926,0.032417663409026)
(1837.83227539,0.0327059918960717)
(1647.30444336,0.0135455111336513)
(1739.10449219,0.025980385898817)
(1637.4041748,0.0123686402754006)
(1628.15588379,0.011787117210418)
(1603.12768555,0.00885758611766388)
(1789.2902832,0.031924256955835)
(1568.34191895,0.00602271941477844)
(1741.96362305,0.024934596951327)
}; 
\addplot [ForestGreen, mark=triangle, only marks, mark size=2pt, mark options={solid,thick}] coordinates {
(2013.15661621,0.0119871729910065)
(1886.71411133,0.031522649832342)
(1868.44873047,0.0451566863523258)
(1820.65563965,0.049544010465171)
(1747.40930176,0.0493345707883921)
(1876.80871582,0.0331716457548021)
(1966.92700195,0.019497728083046)
(1993.74035645,0.0151648055486158)
(1880.93579102,0.0311027909938579)
(1885.56103516,0.0301536291738217)
(1730.24450684,0.0443888757606148)
(2031.7364502,0.0110485849876909)
(1800.84069824,0.0523016102374815)
(1768.06152344,0.0497040279091181)
(1873.86962891,0.0426694282139344)
(1857.6895752,0.0459949481384016)
(1793.28295898,0.0577075328128256)
(1723.5,0.0360995686242212)
(1806.69091797,0.0535869121815874)
(1672.93603516,0.0262081153330282)
(1799.16796875,0.0536218082716568)
(1621.82702637,0.0179882559563831)
(1804.87023926,0.0517686351238139)
(1837.83227539,0.048712260831903)
(1647.30444336,0.0221096214205345)
(1739.10449219,0.0496121982400923)
(1637.4041748,0.0209039368200006)
(1628.15588379,0.0195049183756518)
(1603.12768555,0.0155966318776179)
(1789.2902832,0.0590622441555067)
(1568.34191895,0.0122484083662943)
(1741.96362305,0.048534691997862)
};
\addplot [brown, mark=o, only marks, mark size=2pt, mark options={solid,thick}] coordinates {
(2013.15661621,0.0141544162189226)
(1886.71411133,0.0361398908092942)
(1868.44873047,0.0379850098324585)
(1820.65563965,0.0413921232466621)
(1747.40930176,0.0358020147514111)
(1876.80871582,0.0363408483838952)
(1966.92700195,0.0250295601015855)
(1993.74035645,0.0190238415033753)
(1880.93579102,0.0313743312225083)
(1885.56103516,0.034498973465455)
(1730.24450684,0.0323320153286988)
(2031.7364502,0.0121639943587112)
(1800.84069824,0.0397107957399282)
(1768.06152344,0.0390365522095434)
(1873.86962891,0.033639786477239)
(1857.6895752,0.039475771817161)
(1793.28295898,0.0386739580471282)
(1723.5,0.0300860009215491)
(1806.69091797,0.040660914081387)
(1672.93603516,0.0231051953909517)
(1799.16796875,0.0389530463355412)
(1621.82702637,0.0151884877838829)
(1804.87023926,0.0407788979561552)
(1837.83227539,0.0411953399188674)
(1647.30444336,0.0185381827647755)
(1739.10449219,0.0335077628195194)
(1637.4041748,0.0170209045056422)
(1628.15588379,0.0162782041678722)
(1603.12768555,0.012660598791596)
(1789.2902832,0.0393416145275411)
(1568.34191895,0.00862121222916478)
(1741.96362305,0.0317629186631519)
};
\end{axis}
\end{tikzpicture}

\captionsetup{justification=centering}
\caption{Standard deviations of qualifying probabilities for the Round of 16 \\ in the old and new UEFA Champions League designs, 2022/23 season}
\label{Fig_A4}
\end{figure}

\begin{figure}[ht!]
\centering

\begin{tikzpicture}
\begin{axis}[
name = axis1,
xlabel = Elo rating,
x label style = {font=\small},
ylabel = Standard deviation,
y label style = {font=\small},
y tick label style = {/pgf/number format/.cd,fixed,fixed zerofill,precision=2},
width = 0.99\textwidth,
height = 0.6\textwidth,
ymajorgrids = true,
ymin = 0,
legend style = {font=\small,at={(0,-0.15)},anchor=north west,legend columns=1},
legend entries = {{New design, seeding based on UEFA club coefficients ($\sigma_i^n$) $\qquad \qquad \qquad \quad$}, {Old design, seeding based on UEFA club coefficients ($\sigma_i^o$) $\qquad \qquad \qquad \quad \: \:$}, {New design, seeding based on Elo ratings ($\sigma_i^{n, \mathit{Elo}}$) $\qquad \qquad \qquad \qquad \qquad \quad$}, {Old design, seeding based on Elo ratings ($\sigma_i^{o, \mathit{Elo}}$) $\qquad \qquad \qquad \qquad \qquad \quad \:$}, {New design without the play-offs, seeding based on Elo ratings ($\sigma_i^{n, \mathit{Elo}, T16}$)}}
]
\addplot [red, mark=oplus, only marks, mark size=2pt, mark options={solid,thin}] coordinates {
(2084.97290039,0.00556849568053565)
(1917.42382812,0.0383883782359138)
(1921.01159668,0.0299313501988254)
(1770.73168945,0.0592240653866392)
(1868.74890137,0.0519514276496309)
(1942.46520996,0.0340585151387058)
(1839.03637695,0.0623519987389879)
(1729.8692627,0.0488039664180908)
(1876.95678711,0.043526175250439)
(1889.9822998,0.0477379231775882)
(1858.0637207,0.0529537443722636)
(1845.29052734,0.0552586943231407)
(1865.52929688,0.0452494487547927)
(1804.54443359,0.0633986916671401)
(1911.29858398,0.0402568945841773)
(1754.92822266,0.0595231449083445)
(1588.90246582,0.0123350366083011)
(1823.25415039,0.062879827223615)
(1756.6315918,0.0598732259627556)
(1757.99511719,0.0561958216362157)
(1623.14355469,0.0221030655001632)
(1741.05627441,0.054248315353739)
(1772.43310547,0.0591762252059865)
(1648.07592773,0.0274820161598856)
(1831.67810059,0.065438339681721)
(1585.31518555,0.0125151614158694)
(1737.41174316,0.0532534529044479)
(1624.71594238,0.0202092235555134)
(1627.39501953,0.0208128257984762)
(1801.06494141,0.066963073325447)
(1730.19213867,0.0532501387160003)
(1691.45593262,0.042091906877981)
};
\addplot [blue, mark=asterisk, only marks, mark size=2.5pt, mark options={solid,semithick}] coordinates {
(2084.97290039,0.0127643999038304)
(1917.42382812,0.0714801908831299)
(1921.01159668,0.0602865610683297)
(1770.73168945,0.106768094602391)
(1868.74890137,0.0936214669564362)
(1942.46520996,0.061485664076164)
(1839.03637695,0.103377557135576)
(1729.8692627,0.124520294823064)
(1876.95678711,0.0704673140778996)
(1889.9822998,0.0709044198958059)
(1858.0637207,0.0967233472430618)
(1845.29052734,0.112660982633776)
(1865.52929688,0.0925274836750424)
(1804.54443359,0.114332801896873)
(1911.29858398,0.0746649082201239)
(1754.92822266,0.116708273644747)
(1588.90246582,0.0645536497240295)
(1823.25415039,0.111631757489884)
(1756.6315918,0.12134207853842)
(1757.99511719,0.129555001730823)
(1623.14355469,0.0698935825944549)
(1741.05627441,0.115437825978605)
(1772.43310547,0.119111873944265)
(1648.07592773,0.0824982712592256)
(1831.67810059,0.104482548873097)
(1585.31518555,0.0602544271817017)
(1737.41174316,0.110624470970837)
(1624.71594238,0.0733543238795903)
(1627.39501953,0.0723156438908716)
(1801.06494141,0.116321340493713)
(1730.19213867,0.112894530066425)
(1691.45593262,0.0915001319843483)
};
\addplot [orange, mark=otimes, only marks, mark size=2pt, mark options={solid,thin}] coordinates {
(2084.97290039,0.00452128311444893)
(1858.0637207,0.0294928165190479)
(1889.9822998,0.0274994250372145)
(1911.29858398,0.0237850126513909)
(1624.71594238,0.0119658388882811)
(1845.29052734,0.0318906341822885)
(1865.52929688,0.0228392097170743)
(1876.95678711,0.0218398647131957)
(1942.46520996,0.0196946764451788)
(1921.01159668,0.0164561269779713)
(1804.54443359,0.0308129195632613)
(1648.07592773,0.0145445288104606)
(1729.8692627,0.0261560944576456)
(1627.39501953,0.0121947623276646)
(1868.74890137,0.0277751610359071)
(1917.42382812,0.0231756887394646)
(1754.92822266,0.0292061464419733)
(1839.03637695,0.0324769123198902)
(1756.6315918,0.0284835056496579)
(1772.43310547,0.029016053062664)
(1823.25415039,0.031659789849165)
(1730.19213867,0.0256479849136366)
(1831.67810059,0.0314432228033365)
(1757.99511719,0.0301871261730325)
(1691.45593262,0.0202031202061975)
(1801.06494141,0.0314831489477731)
(1585.31518555,0.00782827393956502)
(1741.05627441,0.0256823919496658)
(1588.90246582,0.00860602246235871)
(1737.41174316,0.0256512414941303)
(1770.73168945,0.0292333077547219)
(1623.14355469,0.0114876332077899)
}; 
\addplot [ForestGreen, mark=triangle, only marks, mark size=2pt, mark options={solid,thick}] coordinates {
(2084.97290039,0.0068611155001141)
(1858.0637207,0.0327241963847705)
(1889.9822998,0.0228212792320191)
(1911.29858398,0.0205667798021818)
(1624.71594238,0.0155287606204232)
(1845.29052734,0.0417070425275684)
(1865.52929688,0.0252936774930242)
(1876.95678711,0.0261714672247602)
(1942.46520996,0.0167323600209855)
(1921.01159668,0.0194480979150279)
(1804.54443359,0.0427613629947078)
(1648.07592773,0.0188373730287061)
(1729.8692627,0.0283472645312807)
(1627.39501953,0.0153097947018833)
(1868.74890137,0.0324466340791008)
(1917.42382812,0.0170924406252158)
(1754.92822266,0.0450101798629076)
(1839.03637695,0.0402750006291207)
(1756.6315918,0.0428809548318049)
(1772.43310547,0.0449183037637555)
(1823.25415039,0.0410050059201375)
(1730.19213867,0.0412070916153931)
(1831.67810059,0.0416880380506989)
(1757.99511719,0.0424640245900502)
(1691.45593262,0.0249054061250616)
(1801.06494141,0.041306949719086)
(1585.31518555,0.0115462372483244)
(1741.05627441,0.0407313442130431)
(1588.90246582,0.0124413786978736)
(1737.41174316,0.040181474710492)
(1770.73168945,0.0413829692765774)
(1623.14355469,0.015656235873342)
};
\addplot [brown, mark=o, only marks, mark size=2pt, mark options={solid,thick}] coordinates {
(2084.97290039,0.00690359033833553)
(1858.0637207,0.0365027060304787)
(1889.9822998,0.0352067150382393)
(1911.29858398,0.0324050009116507)
(1624.71594238,0.0168159869295982)
(1845.29052734,0.0394714754395294)
(1865.52929688,0.0276053578093054)
(1876.95678711,0.0276949286526085)
(1942.46520996,0.0279167136658651)
(1921.01159668,0.0211907444808258)
(1804.54443359,0.0390191245811984)
(1648.07592773,0.0201929322663528)
(1729.8692627,0.0326951270885332)
(1627.39501953,0.0169349806190786)
(1868.74890137,0.0354515411621381)
(1917.42382812,0.0306671034699274)
(1754.92822266,0.0358380100865961)
(1839.03637695,0.0406300771247996)
(1756.6315918,0.0343652336713325)
(1772.43310547,0.0368497678335887)
(1823.25415039,0.0401667739397718)
(1730.19213867,0.0332018836224103)
(1831.67810059,0.0396846979357315)
(1757.99511719,0.0368631233737032)
(1691.45593262,0.0264658074307768)
(1801.06494141,0.03956922171723)
(1585.31518555,0.0117043783035013)
(1741.05627441,0.0317544618257946)
(1588.90246582,0.0118619654874403)
(1737.41174316,0.0327392715848093)
(1770.73168945,0.0357979111323753)
(1623.14355469,0.0164459727903527)
};
\end{axis}
\end{tikzpicture}

\captionsetup{justification=centering}
\caption{Standard deviations of qualifying probabilities for the Round of 16 \\ in the old and new UEFA Champions League designs, 2023/24 season}
\label{Fig_A5}
\end{figure}

\begin{figure}[ht!]
\centering

\begin{tikzpicture}
\begin{axis}[
name = axis1,
xlabel = Elo rating,
x label style = {font=\small},
xmax = 2060,
ylabel = Standard deviation,
y label style = {font=\small},
y tick label style = {/pgf/number format/.cd,fixed,fixed zerofill,precision=2},
width = 0.99\textwidth,
height = 0.6\textwidth,
ymajorgrids = true,
ymin = 0,
legend style = {font=\small,at={(0,-0.15)},anchor=north west,legend columns=1},
legend entries = {{New design, seeding based on UEFA club coefficients ($\sigma_i^n$) $\qquad \qquad \qquad \quad$}, {Old design, seeding based on UEFA club coefficients ($\sigma_i^o$) $\qquad \qquad \qquad \quad \: \:$}, {New design, seeding based on Elo ratings ($\sigma_i^{n, \mathit{Elo}}$) $\qquad \qquad \qquad \qquad \qquad \quad$}, {Old design, seeding based on Elo ratings ($\sigma_i^{o, \mathit{Elo}}$) $\qquad \qquad \qquad \qquad \qquad \quad \:$}, {New design without the play-offs, seeding based on Elo ratings ($\sigma_i^{n, \mathit{Elo}, T16}$)}}
]
\addplot [red, mark=oplus, only marks, mark size=2pt, mark options={solid,thin}] coordinates {
(1765.388794,0.0595302547132651)
(1808.076904,0.0565434048818749)
(1811.814575,0.0551474854925966)
(1753.690796,0.0515790534786919)
(1938.54126,0.035328657844094)
(1942.084595,0.0318247400315867)
(2002.189453,0.0156530907571077)
(1657.92981,0.0292494163934831)
(1843.68335,0.0559533124056938)
(2003.343384,0.0154808907036054)
(1952.873047,0.030719930747727)
(1833.2052,0.0582336431643787)
(1806.794922,0.0594287868352089)
(1745.786377,0.0536501437269456)
(1971.146851,0.0253361776299602)
(1924.533813,0.0384365398884785)
(1733.955566,0.0493217862876296)
(1834.996338,0.0591213927701562)
(1948.962646,0.0261763094070607)
(1802.722778,0.0559780985968845)
(1796.376099,0.0570293880565284)
(1810.3479,0.0605153306298597)
(1909.530151,0.0365961754156777)
(1753.286377,0.0531807223082314)
(1666.642334,0.0293258671736316)
(1721.973145,0.0450610971623803)
(1524.164185,0.00645531618955316)
(1655.044678,0.0288200205514254)
(1685.976196,0.0367704334455793)
(1674.89563,0.0335574486270687)
(1836.174683,0.0572010477514518)
(1523.598145,0.00632975835564367)
};
\addplot [blue, mark=asterisk, only marks, mark size=2.5pt, mark options={solid,semithick}] coordinates {
(1765.388794,0.108884293198355)
(1808.076904,0.0971739428353412)
(1811.814575,0.107494763198563)
(1753.690796,0.109037934551105)
(1938.54126,0.0565355947388234)
(1942.084595,0.0551120524869953)
(2002.189453,0.0374894299417497)
(1657.92981,0.0527130667235687)
(1843.68335,0.106642428758968)
(2003.343384,0.0304494198165937)
(1952.873047,0.0541635897855604)
(1833.2052,0.0980375455442268)
(1806.794922,0.097432204812036)
(1745.786377,0.0840754977405446)
(1971.146851,0.0488871794575563)
(1924.533813,0.071559982323014)
(1733.955566,0.0712273243336553)
(1834.996338,0.0932862455167031)
(1948.962646,0.0484473486332285)
(1802.722778,0.0842122977128999)
(1796.376099,0.0945823810379227)
(1810.3479,0.0972235066290105)
(1909.530151,0.0606498269784219)
(1753.286377,0.0876243739487473)
(1666.642334,0.0639871173767829)
(1721.973145,0.0765773150821128)
(1524.164185,0.0183873324327917)
(1655.044678,0.0596534146036409)
(1685.976196,0.0688932608125453)
(1674.89563,0.063506947244757)
(1836.174683,0.0969335361287642)
(1523.598145,0.0184598679769311)
};
\addplot [orange, mark=otimes, only marks, mark size=2pt, mark options={solid,thin}] coordinates {
(2003.343384,0.0118709180224065)
(2002.189453,0.0120327893282904)
(1938.54126,0.0206666182310776)
(1765.388794,0.0362567249086442)
(1948.962646,0.018658113629309)
(1811.814575,0.0382666982661974)
(1942.084595,0.0216713207290682)
(1806.794922,0.0382213210182534)
(1753.286377,0.0361339976308294)
(1971.146851,0.0170994570215089)
(1721.973145,0.0300013930273843)
(1834.996338,0.0355103489828653)
(1952.873047,0.0196847932955133)
(1745.786377,0.033889712179238)
(1924.533813,0.0243955679322031)
(1796.376099,0.0401591852402306)
(1753.690796,0.0354034739109733)
(1836.174683,0.0387897676852682)
(1666.642334,0.0218813873214454)
(1909.530151,0.0259584825792614)
(1802.722778,0.0360665093555933)
(1810.3479,0.0353994015292438)
(1655.044678,0.0211917868012074)
(1733.955566,0.0311124455526766)
(1808.076904,0.0394710679384813)
(1843.68335,0.0352721714256023)
(1833.2052,0.0372130190092632)
(1685.976196,0.0250742472843529)
(1674.89563,0.0242821445004666)
(1657.92981,0.0212838671063084)
(1524.164185,0.00576642789694718)
(1523.598145,0.00544685452570786)
}; 
\addplot [ForestGreen, mark=triangle, only marks, mark size=2pt, mark options={solid,thick}] coordinates {
(2003.343384,0.0125767473765904)
(2002.189453,0.0123893704642528)
(1938.54126,0.0267191024256885)
(1765.388794,0.0466423707561028)
(1948.962646,0.0188631698071977)
(1811.814575,0.0487322304669742)
(1942.084595,0.0238509511733153)
(1806.794922,0.0507899718176456)
(1753.286377,0.0448625347257823)
(1971.146851,0.0200930538946875)
(1721.973145,0.0291368857139728)
(1834.996338,0.0460452292766486)
(1952.873047,0.0225868089587471)
(1745.786377,0.0440165727274929)
(1924.533813,0.0288139575231891)
(1796.376099,0.0479026130428846)
(1753.690796,0.0482005627067283)
(1836.174683,0.0437173798998214)
(1666.642334,0.0213953152702996)
(1909.530151,0.0309180766456043)
(1802.722778,0.0351398906075689)
(1810.3479,0.0482066078566107)
(1655.044678,0.0197620507450429)
(1733.955566,0.0440549515691228)
(1808.076904,0.0482160954750886)
(1843.68335,0.0443197906114169)
(1833.2052,0.0476284308625261)
(1685.976196,0.0242857318764079)
(1674.89563,0.0230684579838739)
(1657.92981,0.020603701038806)
(1524.164185,0.00824652078789019)
(1523.598145,0.00782380898201042)
};
\addplot [brown, mark=o, only marks, mark size=2pt, mark options={solid,thick}] coordinates {
(2003.343384,0.0164514050381032)
(2002.189453,0.0168533929181294)
(1938.54126,0.027321606727903)
(1765.388794,0.047421403453562)
(1948.962646,0.0254630077713683)
(1811.814575,0.0484350937782648)
(1942.084595,0.0291069292838034)
(1806.794922,0.0484996128690575)
(1753.286377,0.0474069291130308)
(1971.146851,0.0240324996627492)
(1721.973145,0.0417961407474587)
(1834.996338,0.0443859657206301)
(1952.873047,0.0267204532450634)
(1745.786377,0.0436763276454946)
(1924.533813,0.0323793937735274)
(1796.376099,0.0508340611969748)
(1753.690796,0.0459474171215329)
(1836.174683,0.0483720306928844)
(1666.642334,0.0323003633024378)
(1909.530151,0.0338776518562798)
(1802.722778,0.046211291460621)
(1810.3479,0.0442293288387867)
(1655.044678,0.0318771947722278)
(1733.955566,0.0415808651226059)
(1808.076904,0.0495221623979463)
(1843.68335,0.0449316314805896)
(1833.2052,0.0479685615679572)
(1685.976196,0.0352756149203385)
(1674.89563,0.034578331611242)
(1657.92981,0.0317136320603374)
(1524.164185,0.0108822147837009)
(1523.598145,0.0104130755852026)
};
\end{axis}
\end{tikzpicture}

\captionsetup{justification=centering}
\caption{Standard deviations of qualifying probabilities for the Round of 16 \\ in the old and new UEFA Champions League designs, 2025/26 season}
\label{Fig_A6}
\end{figure}

\input{figures/Figure_A7_standard_deviation_number_of_draws}

\end{document}